# Zirah: A Post-Quantum True Random Number Generator Hardware Signer With an Avalanche Noise

Muhammad Shaheer Bin Junaid
Independent Researcher
Karachi, Pakistan
*contact@mshaheerbjunaid.com*
ORCID: 0009-0002-4088-6232

***Abstract-*** Zirah is a hardware signer built around an STM32F407 microcontroller, and its purpose is to find out, on inexpensive hardware, where a post-quantum signing key is exposed and what actually protects it. The device draws every random bit from a reverse-biased 2N3904 junction in avalanche breakdown, amplified by an LM358 and screened by SP 800-90B health tests, computes ML-DSA-65 (FIPS 204, hedged) on the STM32 with two-share arithmetic masking of the secret-key multiplications, and adds ECDSA P-256 from an ATECC608 secure element over the same digest, once a person has matched an eight-digit digest prefix on its OLED with the PC and pressed a button. A desktop application and a command-line tool drive the board and verify its output, and because only a 32-byte manifest digest reaches the device, it signs files of any type and size, returning signed copies for PDF, Office Open XML, ZIP, PNG and JPEG files and detached records for every other format. NIST's SP 800-90B tool rejected the IID assumption for 1,000,000 raw samples and assessed 6.898 bits of min-entropy per byte on the non-IID track, against the 1 bit the firmware credits, while spectral lines near 116 Hz that hold about 5 % of the variance remain unexplained. At 168 MHz the masked signature took a median of 157 ms over 20 runs and a complete signing operation under 0.8 s, and the signer matched the pq-crystals reference in 1,976 checks natively and on an emulated Cortex-M4. The ATECC608 can neither store the 4,032-byte ML-DSA-65 key nor compute with it, so that key sits in ordinary flash behind readout protection, and on an earlier board, before that protection was enabled, a debugger copied it out and the copy produced a valid ML-DSA-65 signature without the device, although the leakage results come from an instruction-level emulator and nothing here is certified.

## 1. Introduction

A signature on a contract, a firmware release or a certificate is often checked years after it was made, which is why the timetable for retiring classical signatures matters to anyone who signs files today. NIST's report on the transition to post-quantum cryptography, published in final form in October 2026, deprecates ECDSA at every security strength after 2030 and disallows it after 2035, because a cryptographically relevant quantum computer would break it [1]. The post-quantum replacement, ML-DSA, was standardised in FIPS 204 in August 2024 [2], and while confidence in it builds, many systems are expected to carry a classical and a post-quantum signature side by side. The same NIST report accepts such dual signatures as long as at least one component algorithm is approved and every component signature verifies, and the IETF's composite ML-DSA specification, now waiting in the RFC Editor queue, defines one way to package them [3]. The migration therefore has a shape, and hardware that signs has to fit into it.

The inexpensive hardware that usually protects signing keys has not caught up with that shape. The ATECC608, a widely sold secure element, implements ECDSA and ECDH on P-256, SHA-256 with HMAC, and AES-128 [4]. It has no lattice arithmetic, and the largest slot in its data zone holds 416 bytes [5], roughly a tenth of the 4,032 bytes of an ML-DSA-65 secret key [2]. A device built around such a chip can keep its classical key inside the chip, but its post-quantum key has to be stored and used by an ordinary microcontroller, so the half of the signature that is meant to survive a quantum attacker ends up with the weaker protection.

Zirah is a piece of hardware built to study that constraint, and this paper reports what I built and what I measured on it. It began as a true random number generator, a reverse-biased transistor junction whose noise an STM32F407 samples, tests and conditions, and it grew into a signing device once that randomness had a demanding use: keys, masks and hedging values for ML-DSA-65 signatures that the same STM32F407 computes with masked secret-key arithmetic, joined by an ATECC608 that adds an ECDSA P-256 signature over the same digest. A desktop application and a command-line tool are the device's interface to a PC. They send it only a digest of a manifest that describes a file, so a firmware image, a spreadsheet, a photograph or a compiled program is signed by the same path as a PDF, and they check every signature it returns. A person has to compare a digest prefix on the board's OLED with the one on the PC and press a button before anything is signed, and my aim throughout was a prototype whose properties are stated exactly, including the ones that fall short of a product.

The consequence of the constraint was not hypothetical for me. On the STM32F4DISCOVERY board I used before the board measured here, with readout protection still at its factory level, a debugger copied the ML-DSA-65 key out of flash, and the copy produced a valid post-quantum signature with no device involved. The ATECC608 on the same build kept its classical key throughout. That asymmetry, a secure element sitting beside a post-quantum key it cannot protect, is the thread that runs through the rest of this paper.

The contributions are the following, listed in the order the sections address them. I describe the device itself, with its wiring, its confirmation step and its trust model, and an avalanche-noise entropy source built from discrete parts, with SP 800-90B health tests on the device, a SHAKE256 conditioner and the faults found while building it (Sections 4 and 5). I give a masked ML-DSA-65 signer for the STM32F407, derived from the pq-crystals reference code, whose secret vectors exist in RAM only as two arithmetic shares refreshed before every signature in two separate passes over coefficients visited in random order, and whose signatures stay byte-identical to the reference (Section 6). I set out what the secure element protects in such a device and what it leaves to the microcontroller's readout protection, including the key extraction on the earlier board (Section 7). I describe the host software that extends the device, with a signed manifest that binds a file's content, name, size and times, signed copies of PDF, Office Open XML, ZIP, PNG and JPEG files that open like the originals, detached records for every other format, and a verification policy that rejects a signature stripped of its ECDSA half (Section 8). Finally, I report the measurements on the finished device, among them an SP 800-90B assessment of 1,000,000 raw samples with NIST's own tool (Sections 9 and 10), and the source code, the circuit drawings and the raw measurement data behind them are released in an open archive [6].

I make no claim that Zirah resists side-channel or fault attacks, and the evidence would not support one. Only the secret-key products are masked, the leakage evaluation ran in an instruction-level emulator rather than on the chip, and the entropy results describe one board measured in one session. The device is a single prototype with no certification or independent audit, and the places where it falls short are as much a part of the result as the places where it works.

## 2. Background

### 2.1 Entropy sources and NIST SP 800-90B

NIST SP 800-90B describes how an entropy source is built from a noise source, health tests and an optional conditioning component, and how its min-entropy is assessed from raw samples [7]. The assessment first asks whether the samples look independent and identically distributed (IID), using chi-square tests, a longest-repeated-substring test and 19 permutation tests that compare statistics of the data with those of shuffled copies. When the IID assumption fails, ten non-IID estimators are run on the raw symbols and on their bits, and the lowest estimate, bounded by the bit-string results, becomes the assessed min-entropy. NIST publishes an implementation of these estimators [8], and an assessment made with it is the evidence that SP 800-90B asks for, although running the tool is not the same as a validation.

A second NIST document, SP 800-22, is often meant when people speak of the NIST randomness tests [9]. Its statistical test suite looks for patterns in the output of a random or pseudorandom generator, which here would be the conditioned bytes that leave SHAKE256. Those bytes say very little about the noise source, because a conditioner makes even a weak source look uniform, and the Zirah firmware has no command that returns its conditioned output in any case. SP 800-90B, applied to the raw samples, is therefore the assessment that fits this design, and it is the one reported in Section 10.3.

## 2.2 ML-DSA and its implementations on the Cortex-M4

ML-DSA is the standardised form of CRYSTALS-Dilithium, a signature scheme whose security rests on the hardness of module lattice problems [10, 2]. FIPS 204 defines three parameter sets, and ML-DSA-65, the one Zirah uses, targets NIST security category 3 with a 1,952-byte public key, a 4,032-byte secret key and a 3,309-byte signature [2]. The standard offers a pure mode, which signs the message itself under an optional context string, and a separate HashML-DSA mode that signs a hash together with an identifier of the hash function. It also offers a hedged variant of signing that mixes fresh randomness into each signature and a deterministic variant that does not, and the difference matters for fault attacks, as Section 2.3 explains. A verifier has to use the same mode and context as the signer, so these choices are part of a signature format rather than implementation details.

On microcontrollers the transform that dominates ML-DSA's running time is the number-theoretic transform (NTT). Greconici, Kannwischer and Sprenkels gave fast and compact Dilithium implementations for the Cortex-M3 and Cortex-M4 [11], and their assembly NTT is part of pqm4, the library and benchmark suite of post-quantum implementations for the Cortex-M4 [12]. Zirah uses that transform unchanged and otherwise builds on the pq-crystals reference implementation [13], which keeps the masked parts of the code close to a well-reviewed original.

## 2.3 Side-channel attacks on the secret-key products

The multiplication of the challenge polynomial with the secret vectors is the most obvious target for power analysis in ML-DSA, because it mixes a public value with every secret coefficient. Chen and colleagues attacked the reference implementation running on an STM32F405, a part from the same family as Zirah's STM32F407, with a non-profiled correlation attack on the NTT-domain pointwise product of the challenge and the secret polynomials, and recovered a secret coefficient with 99.99 % confidence from 157 traces [14]. Masking is the standard countermeasure. Migliore and colleagues presented a masked Dilithium at arbitrary order and evaluated it [15], Azouaoui and colleagues revisited which intermediate values are sensitive and gave improved masked implementations [16], and Coron and colleagues designed faster gadgets for high-order masking [17]. Schemes of that kind also protect the masking vector $\boldsymbol{y}$, its SHAKE-based generation and the rejection checks, which needs masked Keccak and conversions between Boolean and arithmetic masking, so a complete masked ML-DSA is a far larger engineering task than masking two products.

Two further results shaped the details of Zirah's signer. Balasch and colleagues showed that masked software loses security order when the shares of one value pass through the same registers in consecutive instructions, because the leakage then depends on the distance between the shares [18]. Veyrat-Charvillon and colleagues analysed what shuffling adds to masking and where it stops helping [19]. For fault attacks, Groot Bruinderink and Pessl showed that deterministic lattice signatures, Dilithium among them, are open to differential fault attacks that rely on the same internal randomness being reused when the same message is signed twice [20]. The hedged variant of FIPS 204 draws fresh randomness for every signature and removes that precondition, which is one reason Zirah uses it.

## 2.4 Secure elements and readout protection

The ATECC608 holds up to 16 keys or data items in a data zone of 1,208 bytes and computes ECDSA and ECDH on P-256 internally, so a private key generated inside it never has to leave [4, 5]. A key that does not fit those algorithms gets no such protection and has to live in the microcontroller, where the STM32F407 offers readout protection (RDP) in three levels [21]. At level 0 a debugger reads the flash freely, at level 1 the flash is closed to the debug interface and to the system bootloader and returning to level 0 erases it, and level 2 disables the debug interface permanently. Readout protection is weaker than its name suggests, and Obermaier and Tatschner extracted firmware from read-protected STM32F0 devices [22], so level 1 raises the cost of reading a key without ruling it out.

## 2.5 Hybrid signatures in the transition

A hybrid or dual signature attaches two or more component signatures to one message and accepts the result only when every component verifies [1]. Read that way, it is an AND combination: a forger has to break every component algorithm, or steal every key, before a forged signature passes. NIST's transition report states that verification of a dual signature requires all component signatures to verify and that existing standards accept such signatures when at least one component algorithm is approved [1], and the composite ML-DSA draft gives a concrete encoding for certificates [3]. Zirah does not use the composite encoding, because it signs files rather than certificates, but it follows the same rule, so its security against a quantum attacker rests on the ML-DSA half and its security against a cryptanalytic surprise in ML-DSA rests on the ECDSA half for as long as ECDSA holds.

## 3. Threat Model

Zirah was designed against the attackers in Table 1, and each row also states what that attacker still has, because a prototype is only useful to others when its residual risks are written next to its defences. The device is assumed to be in the hands of its owner, who signs files on a Windows PC that may or may not be trustworthy, and the people who verify a signature are assumed to hold the owner's exported public keys or a published fingerprint. The design goal is that no file is signed without a deliberate press of the board's button and that no signature is released unless it verifies, while the honest limits are those listed in the right-hand column.

**Table 1. Attackers Zirah was designed against, what the design does about each, and what remains.** Source: docs/DESIGN.md in the Zirah repository [23].

| Attacker | What Zirah does | What remains |
|---|---|---|
| Malware on the PC | Button press for every signature; the OLED shows 32 bits of the manifest digest; the PC cancels if the device's echoed prefix differs | A matching fake on screen; a 32-bit prefix can be matched with about $2^{32}$ attempts; trust and verification run on the PC |
| Brief physical access with a debugger | ECDSA key inside the ATECC608; readout protection level 1 | Published readout-protection bypasses on other STM32 families; the ML-DSA key is unmasked at rest |
| Power or EM side channel | $c\boldsymbol{s}_1$ and $c\boldsymbol{s}_2$ on two shares, refreshed and shuffled | $\boldsymbol{y}$, the recombined values, the rejection checks and the boot-time key load are unmasked; no physical measurement |
| Fault injection | Hedged signing; every signature verified before release; faults wipe secrets and halt | A fault that corrupts a signature and its check consistently, or one in the boot-time key load |
| Entropy manipulation | Start-up and continuous health tests; a window on the bias voltage | Subtle bias, injected signals below the tests' sensitivity, slow drift |
| Malicious USB host | Bounded printable command lines, strict parsing, no heap, MPU guard region, stack canary | Firmware bugs that pass these checks |

Two rows deserve more than a table cell. A compromised PC can display a harmless file while it sends the digest of another, and the OLED shows only the first 32 bits of the manifest digest, so an attacker who can vary a file's bytes, name or times can find a malicious manifest whose digest starts with the same 32 bits after about $2^{32}$ attempts on average, which is cheap. The OLED check therefore stops careless misuse and naive malware rather than a determined attacker in control of the PC. The entropy row has a similar shape, since the health tests catch a dead, stuck or grossly repeating source, as they did during bring-up, but they cannot detect a subtle bias, a signal injected below their sensitivity or slow drift.

## 4. The Device

The device is the subject of this paper, and the software around it exists to drive it and to check what it returns. It consists of one microcontroller board and two solderless prototyping boards, Board A with the noise source and Board B with the secure element and the display, and it holds both signing keys, the post-quantum one in the STM32's flash and the classical one inside the ATECC608. Everything that touches a key, from the health tests to the final self-check of each signature, runs on the board, while the PC only hashes files, builds the manifest and verifies the result with independent code.

### 4.1 Hardware and wiring

Figure 1 shows the parts of Zirah and what passes between them, and Figure 2 shows the device as built. The main board is a DevEBox STM32F4XX_M core board with an STM32F407VE, a Cortex-M4F running at 168 MHz with 512 KB of flash, 128 KB of main SRAM and 64 KB of core-coupled memory (CCM). It has no on-board debugger, so a CP2102 USB-to-UART adapter on USART1 (PA9 and PA10, 115,200 baud) carries the host protocol and also, through the STM32's ROM bootloader [24], firmware updates and the command that sets readout protection. The board's K1 button on PA0 confirms each signature. Board B holds the ATECC608 module and a 1.3-inch SH1106 OLED on a software I2C bus, and Board A is the noise source whose output the firmware samples on PA4 while it watches the bias supply on PB0 (Section 5).

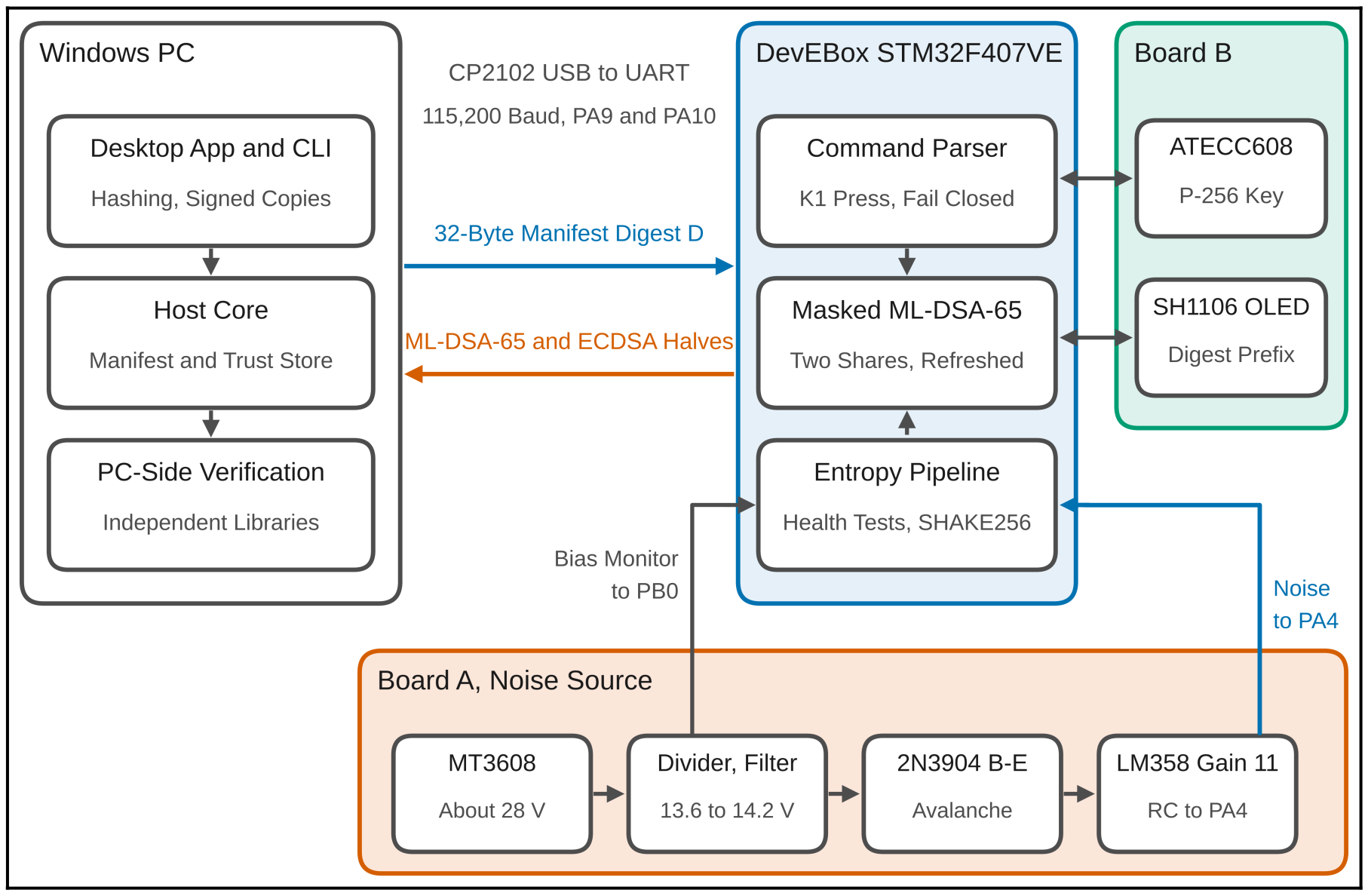

***Fig. 1*** *Zirah's parts and what crosses the serial link. Only the 32-byte manifest digest D travels to the board, and the file itself stays on the PC; the ATECC608 generator, the STM32's own generator and the jitter of its internal LSI oscillator are mixed into the entropy pipeline, credited with zero entropy*

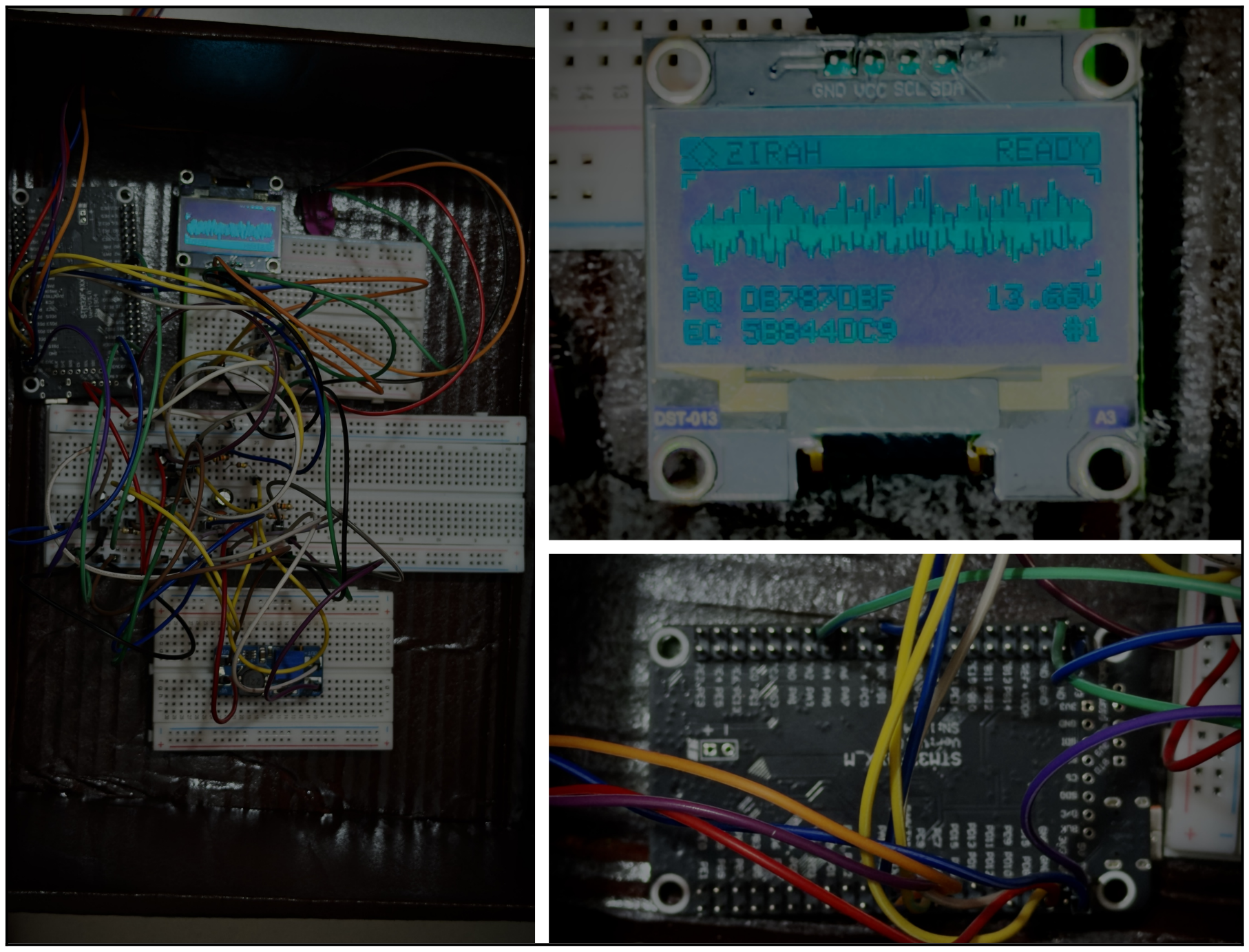

***Fig. 2*** *The device as built. Left, the whole build in its box: the DevEBox at the top left, the OLED on Board B beside it with the ATECC608 module (purple) on jumper wires, Board A on the long board and the MT3608 boost module on the small board below. Top right, the OLED in the Ready state, drawing a live block of noise samples. Bottom right, the DevEBox and its header wiring*

Figure 3 gives every connection between the boards, drawn from the firmware's pin assignments and from the as-built layout data, and Figure 4 gives the I2C bus of Board B as a schematic. The PC powers the DevEBox through its USB-C socket and talks to it through the CP2102, whose own 5 V and 3.3 V pins stay unconnected so that two supplies never meet. The DevEBox feeds Board A with 5 V for the LM358 and the boost module, 3.3 V for the mid-rail reference and two ground wires, one of them to the lower part of a ground rail that is split part-way along that board. The OLED answers at I2C address 0x3C and the ATECC608 at 0x60, the wiring guide specifies the 4.7 kΩ pull-ups RP1 and RP2 on both lines, and the firmware also enables the STM32's weak internal pull-ups on them. Table 2 lists the main parts.

**Table 2. Main parts of the build.** Source: hardware/BOM.md and hardware/layout/zirah_layout.py in the Zirah repository [23].

| **Part** | **Use** |
|---|---|
| DevEBox STM32F4XX_M (STM32F407VE) | Main board, 168 MHz Cortex-M4F |
| CP2102 USB-to-UART adapter | Host link, flashing and readout protection through the ROM bootloader |
| ATECC608 breakout | ECDSA P-256 key, extra random input |
| SH1106 OLED, 1.3 inch, I2C | Digest prefix, fingerprints, live noise |
| MT3608 boost module | Bias supply, halved by R16 and R17 |
| 2N3904 | Avalanche junction (Q1) |
| LM358 | Noise amplifier with a gain of 11 (U1A) |
| Resistors | R1 and R8 1 kΩ; R6, R10, R16 and R17 10 kΩ; R2, R3, R4, R5, R7 and R9 100 kΩ; RP1 and RP2 4.7 kΩ |
| Capacitors | C1 100 µF; C4, C5 and C9 10 µF; C2, C3, C7 and C8 100 nF; C6 100 pF |

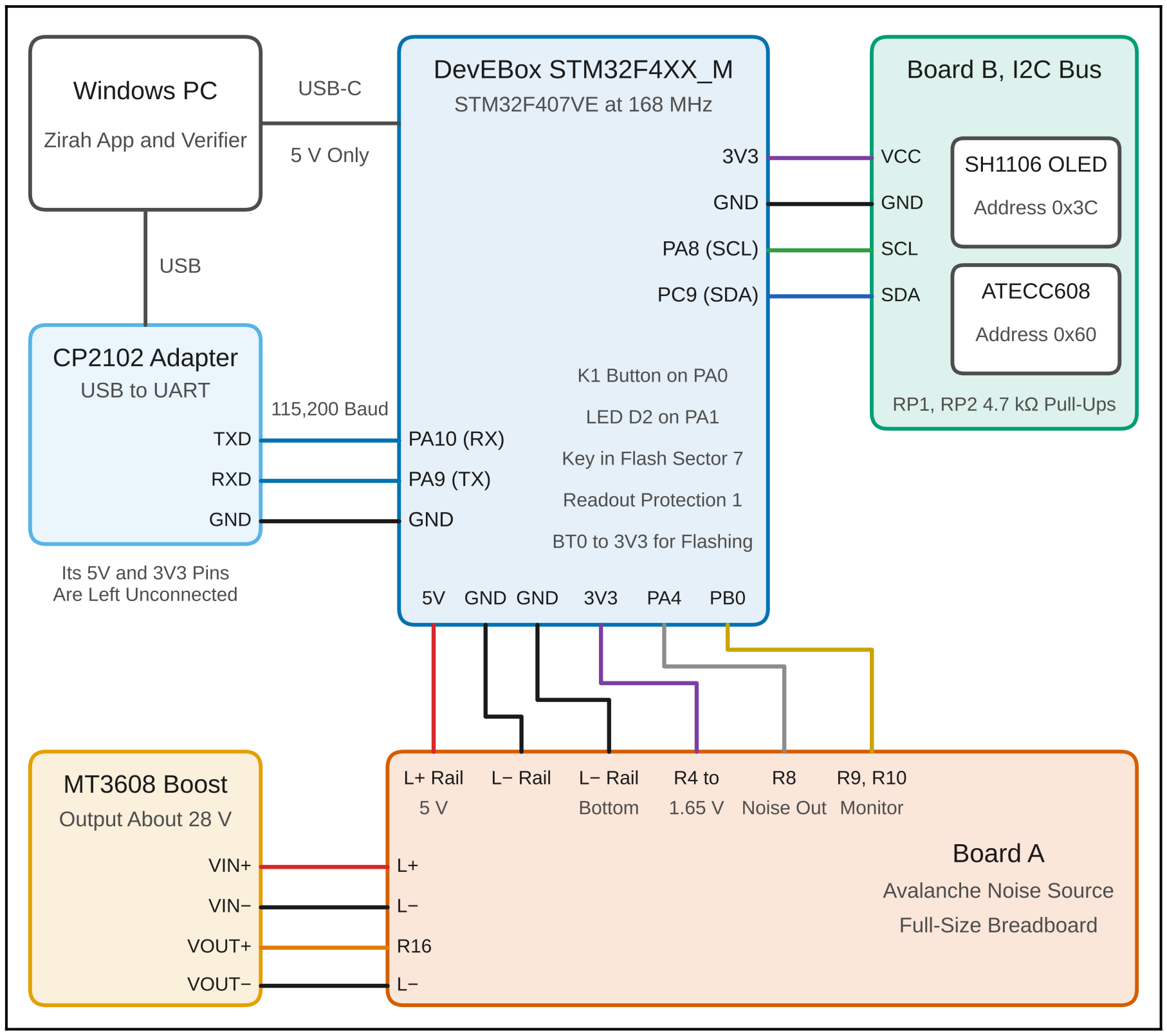

***Fig. 3*** *Every connection of the measured build. The CP2102 carries the serial link on PA9 and PA10, the DevEBox's bottom header feeds Board A and reads its two outputs on PA4 and PB0, and Board B shares 3V3, GND, PA8 and PC9. Wire colours match the layout drawings*

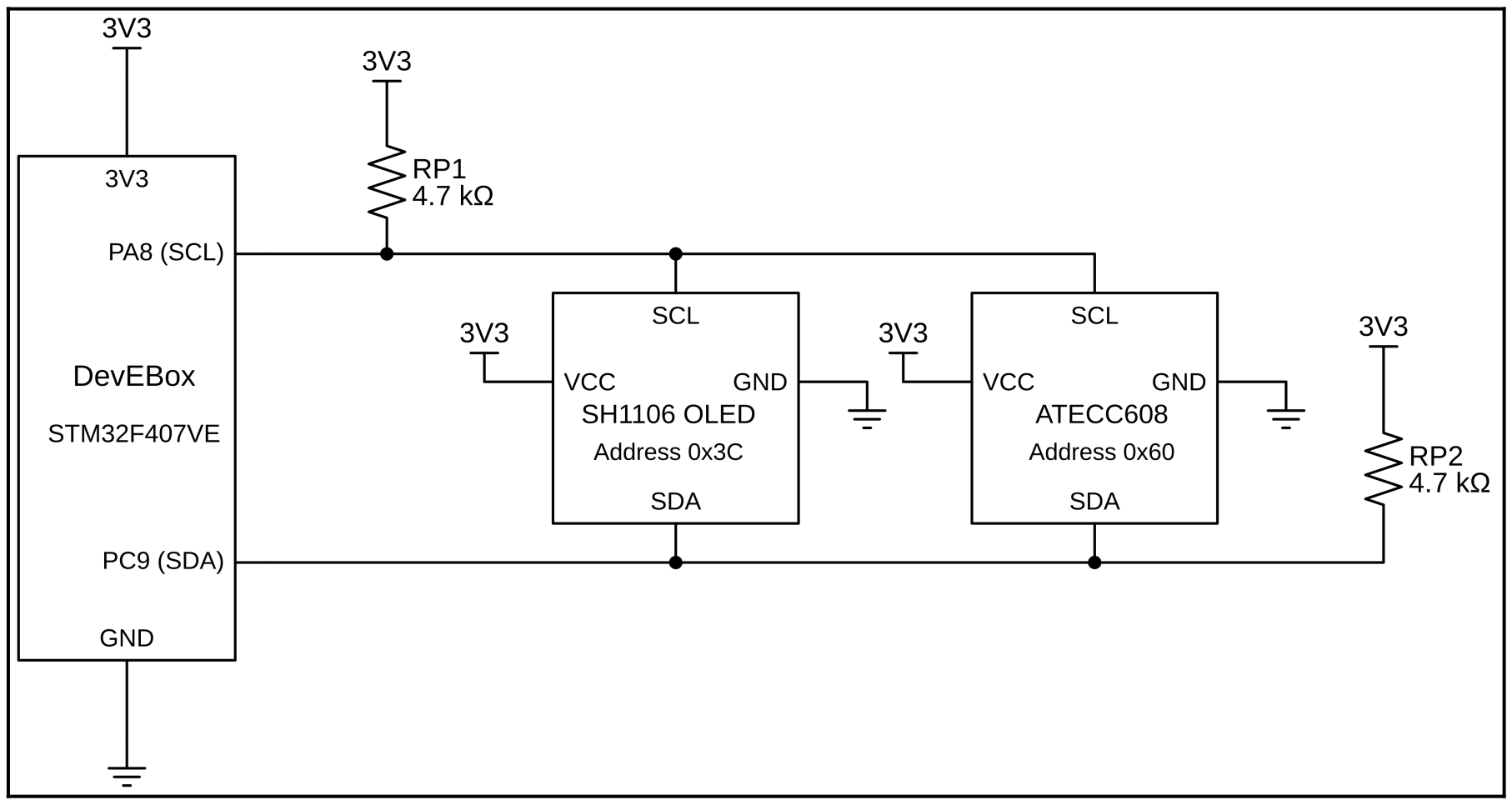

***Fig. 4*** *Board B, the I2C bus of the OLED and the ATECC608, with the pull-ups RP1 and RP2 that hardware/WIRING.md specifies. On the measured build the OLED sat on Board B and the ATECC608 module beside it on jumper wires, with the same four connections*

The firmware was first written for the STM32F4DISCOVERY and builds for either board from one source tree. During the project the ATECC608 module moved from the Discovery to the DevEBox, and its ECDSA key, with the fingerprint 5B844DC9, moved with it, whereas the DevEBox had to generate a new ML-DSA-65 key, fingerprint DB787DBF. That small event is the argument of Section 7 in miniature, because the classical key travelled safely inside its chip while the post-quantum key could not travel at all. All results in Section 10 come from the DevEBox.

### 4.2 Signing a file

A signature passes through six steps, and the device reports each one as it happens so that the desktop application never advances a stage on a timer.

1. The PC hashes the file with SHA-256 and builds the manifest of Section 8.2, then sends only the 32-byte manifest digest $D$ to the board.
2. The board announces the first four bytes of $D$, which the PC compares with its own value, cancelling at once if they differ, and shows them as eight hexadecimal digits on the OLED while the PC shows the same digits on screen (Figure 5).
3. The person compares the two displays and presses K1 within 30 s. A button that is already held down has to be released first, and a press that lasts longer than five seconds is not taken as consent.
4. The board runs a fresh SP 800-90B start-up test on the noise source, refreshes the masks on its key, draws the 32-byte hedging value `rnd` from the conditioner, computes the masked ML-DSA-65 signature and verifies it against its own public key. A signature that fails this check is never released, and signing stays locked until the next power cycle.
5. The ATECC608 signs the same $D$ with its P-256 key.
6. The PC verifies the ML-DSA-65 signature with dilithium-py [25] and the ECDSA signature with pyca/cryptography [26], neither of which shares code with the firmware, hashes the file a second time, and writes the signed copy or the detached record through a temporary file and an atomic rename that never replaces an existing file.

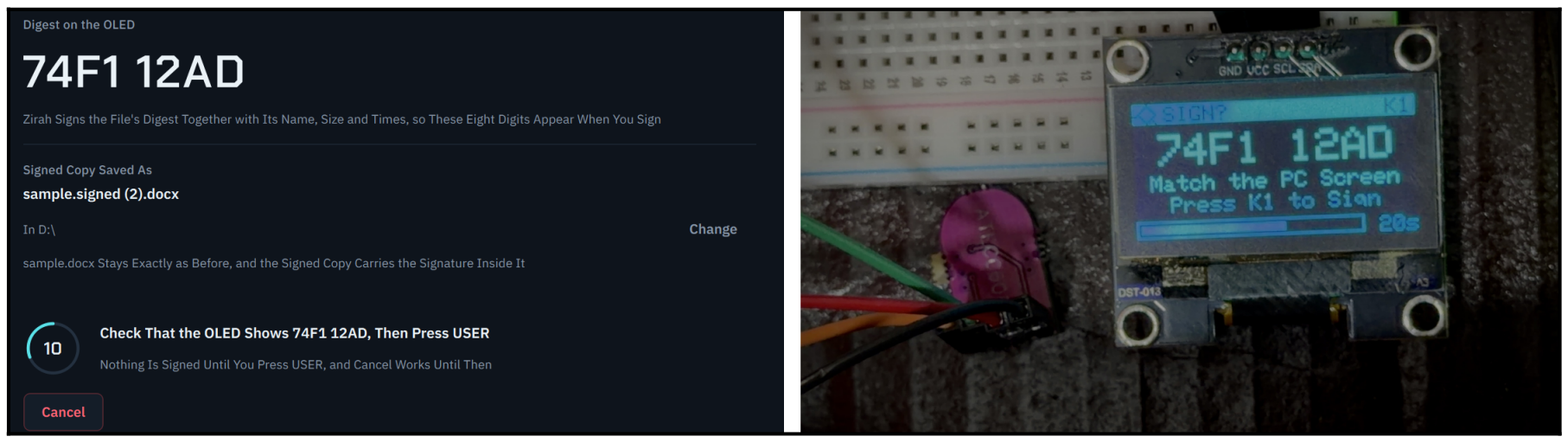

***Fig. 5*** *The confirmation step for one signature. Left, the Sign page of the desktop app, cropped around the digest and the prompt; right, the OLED during the same signature. Both show 74F1 12AD, the first four bytes of the manifest digest, and the OLED counts down the time left to press K1. The prompt in this host build still names the Discovery board's USER button, and the released host code names the board's own button*

### 4.3 Trusting a device

The PC signs only with keys it has pinned in its trust store, which closes the easiest attack on a host tool, a device swapped for one with different keys. A device becomes trusted when the person types the ML-DSA-65 fingerprint the OLED shows, the first four bytes of SHAKE256 of the public key written as eight hexadecimal digits, and the PC never displays the value it expects, so the typed value has to come from the board. If a device's keys change, the tools say so and refuse to sign until the new keys are confirmed in the same way. The fingerprint is there for comparing keys by eye, and at 32 bits it is no barrier against an attacker who controls the PC, a case Section 3 already counts among the residual risks.

## 5. The Avalanche Noise Source

### 5.1 Circuit as built

Figure 6 shows the noise source as it stands on the measured board, and Figure 7 shows its layout on the board. An MT3608 boost module raises the board's 5 V, and a divider of two 10 kΩ resistors, R16 and R17, halves its output to the bias supply, which the firmware measured between 13.6 and 14.2 V during the measurements reported here. R1 (1 kΩ) with C1 (100 µF) and C2 (100 nF) filters that supply, and R2 (100 kΩ) feeds the emitter of a 2N3904 whose base is grounded and whose collector is left open, so the base-emitter junction is reverse-biased into avalanche breakdown with R2 limiting the current to tens of microamps. C3 (100 nF) passes the noise to the non-inverting input of one half of an LM358, held at 1.65 V by R4 and R5 (100 kΩ each, from the 3.3 V rail) through R3. R7 and R6 set a gain of 11 for the noise, and C5 in series with R6 keeps the gain for the DC level at one, so the output stays centred in the ADC range. R8 (1 kΩ) and C6 (100 pF) sit in front of the ADC input on PA4, and R9 (100 kΩ) and R10 (10 kΩ) with C7 divide the bias supply by 11 for the monitor on PB0.

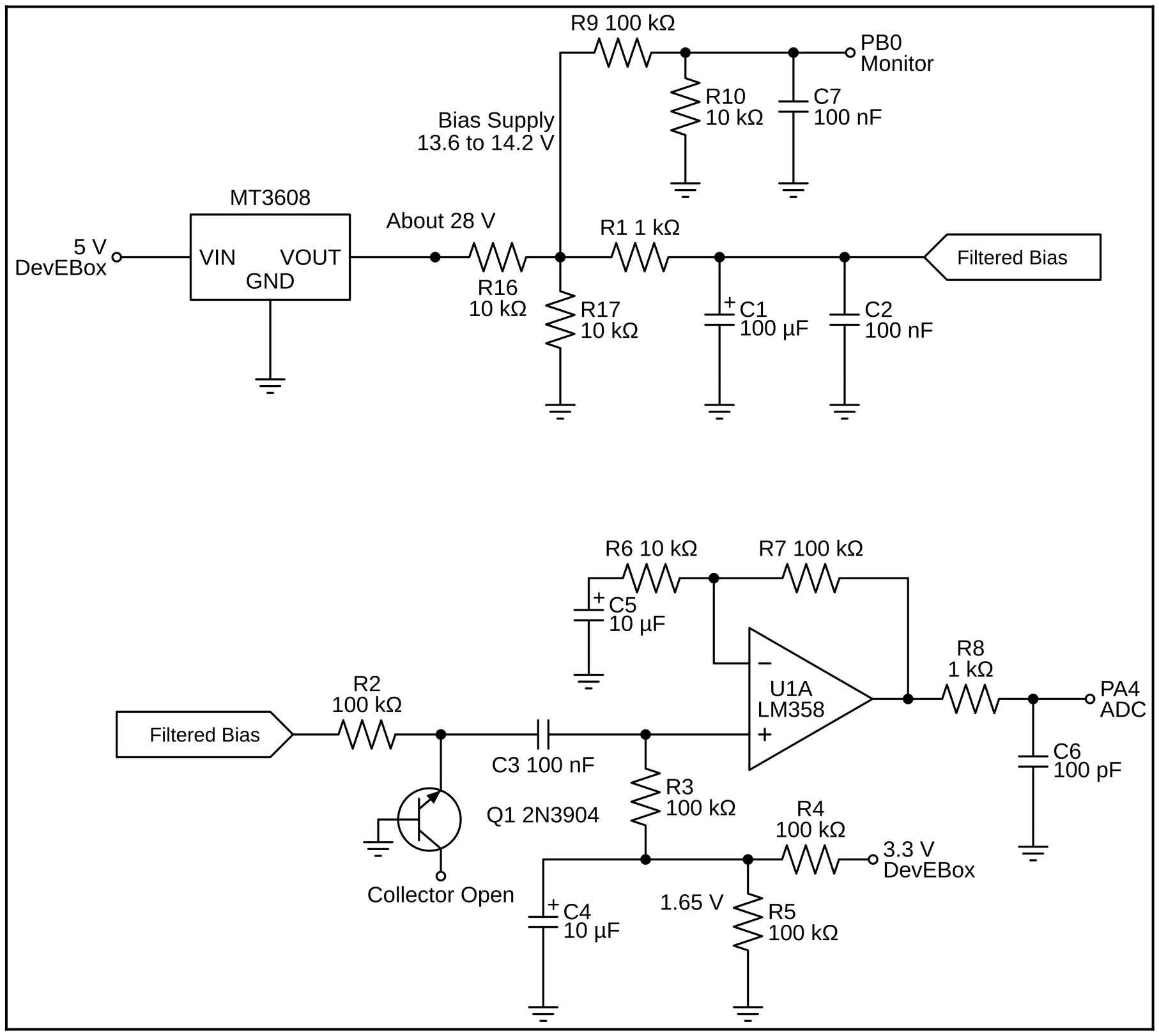

***Fig. 6*** *Board A as built, in two rows joined by the filtered bias net. Top, the bias supply: R16 and R17 halve the boost output because the MT3608 modules would not regulate down to the design value, and R9 and R10 feed a 1/11 copy of the bias to PB0. Bottom, the avalanche junction and the amplifier, whose output reaches PA4 through R8 and C6. The LM358's supply decoupling (C8, C9) and its unused second half, parked as a follower with a grounded input, are not drawn*

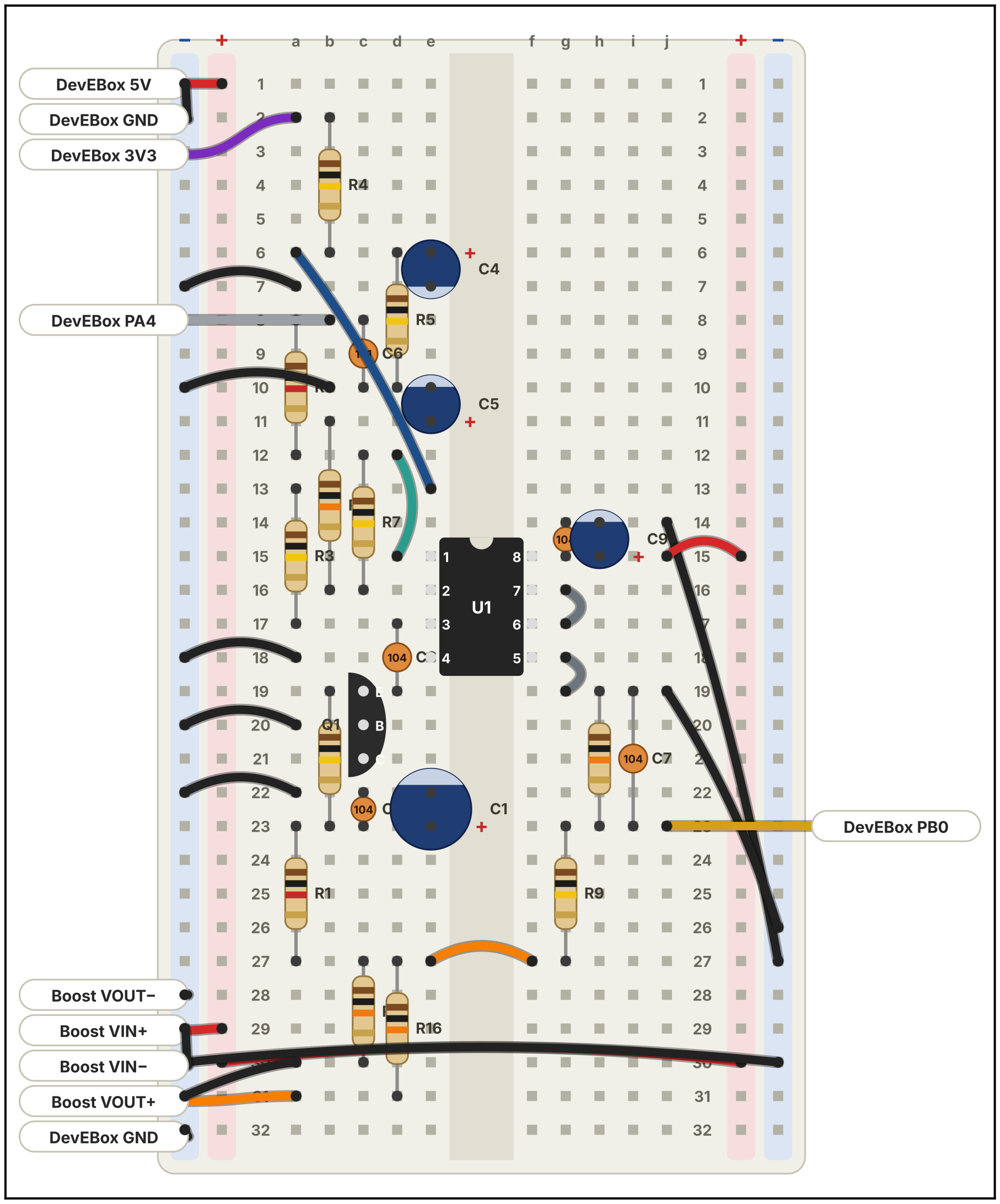

***Fig. 7*** *The layout of Board A as built, generated from the same layout data that a netlist check and a build-order check verify. Some part labels are hidden under neighbouring parts here, and the interactive build guide in the archive places the parts one at a time*

## 5.2 Bring-up

The design called for the MT3608 to make about 12 V, and the board refused. With either of the two modules tried, the output could not be brought below about 26 V on this build, whichever way the trimmer was turned, and the multimeter read between 28 and 30 V at the output during the session. I did not find the cause, and the modules themselves are not necessarily at fault. Rather than wait for other modules, I added the R16 and R17 divider to halve the output, which puts about 5 kΩ of source resistance in front of the existing filter and draws about 1.4 mA from the boost output.

The start-up checks earned their place during bring-up, because they rejected the source for real faults. With the boost output wired straight to the bias network the monitor read 36.4 V and the firmware reported `FAULT_BIAS`, since the window is 9 to 20 V. At a bias of 9.2 V, during adjustment, the samples varied by only 3.8 ADC steps and the source was reported stuck. In both cases the device refused to sign, as intended. Two wiring faults, a ground rail split part-way along the board and a missing ground connection on the monitor divider, also had to be found before the source passed, and the second ground wire in Figure 3 is the repair for the first of them.

### 5.3 Sampling and health tests

ADC1 samples PA4 at 20 kHz, paced by the CPU cycle counter, in blocks of 64 samples for the generator and 2,048 for raw captures. Each 12-bit sample becomes an 8-bit symbol by keeping its low byte, since the low bits of a signal with a standard deviation near 95 ADC steps vary far more than the high bits. The symbol is the unit that the health tests count and that the assessment in Section 10.3 measures.

Every symbol passes through the two continuous health tests of SP 800-90B, Section 4.4 [7], with the assumed min-entropy $H=1$ bit per symbol and a false-alarm probability of about $2^{-20}$ per test. The repetition count test fails when one symbol repeats $C=1+\lceil 20/H \rceil=21$ times in a row, and the adaptive proportion test, with a window of 512 symbols, fails when the window's first symbol occurs 311 times or more. Before every key generation, signature and identity proof the firmware runs a start-up test: it pulls the noise pin to ground for 5 ms and releases it, so that a pin with nothing connected starts near 0 V, then takes 1,024 samples, which must pass both tests, show a standard deviation of at least 8 ADC steps, average within 700 steps of mid-scale and have a lag-1 autocorrelation below 0.4, while the bias monitor must read between 9 and 20 V. Any failure puts the source into a fault state, and the firmware refuses to generate keys, refresh masks or sign until a later start-up test passes.

### 5.4 Conditioning and extra inputs

The output comes in 32-byte blocks, each the SHAKE256 hash of a domain label, a block counter, a header describing the inputs, the additional input, the clock-jitter readings and enough source symbols to carry twice the output size in credited entropy. With $H=1$ that means 512 symbols per block, about 26 ms of sampling. The additional input holds 32 bytes from the ATECC608's generator and 32 bytes from the STM32's own generator, which is used only while its seed-error, clock-error and repeated-word checks pass, and the jitter readings are 256 periods of the internal LSI oscillator timed against an 84 MHz counter, about 8 ms of work. All of these are credited with zero entropy, so the avalanche source alone has to pass its tests and carry the full credited entropy, and a missing or faulty extra input cannot weaken the output. What they add is cover for a weakness in the avalanche source that the health tests miss, because an attacker would also have to predict them.

The conditioner's output goes to the signer directly, with no SP 800-90A deterministic random bit generator in between, so Zirah has an entropy source with conditioning rather than a complete SP 800-90C random bit generator. Whether SHAKE256 used in this way counts as a vetted conditioning component under SP 800-90B Section 3.1.5.1.1 has not been settled for this design, and a formal validation would have to settle it.

## 6. The Masked ML-DSA-65 Signer

### 6.1 Key shares in RAM

The signer begins from the pq-crystals reference implementation of ML-DSA-65 ($k=6$, $\ell=5$) [13], unchanged except that pqm4's Cortex-M4 assembly replaces the C transform [12]. Zirah adds a signer that follows the reference function `crypto_sign_signature_internal` step for step but never holds $\boldsymbol{s}_1$ or $\boldsymbol{s}_2$ whole. At boot the packed secret key is read from flash, $\boldsymbol{s}_1$, $\boldsymbol{s}_2$ and $\boldsymbol{t}_0$ are moved to the NTT domain, $\boldsymbol{s}_1$ and $\boldsymbol{s}_2$ are reduced to the range $0 \le \hat{s} < q$ with $q$ = 8,380,417, and every coefficient $\hat{s}$ is split into two arithmetic shares,

$$\hat{s}^{(1)}=r\,,\ \hat{s}^{(0)}=\hat{s}-r \bmod q\,,\ \hat{s}=\hat{s}^{(0)}+\hat{s}^{(1)} \bmod q\,,$$

where $r$ is drawn uniformly from the integers $0$ to $q-1$ by rejection sampling of 23-bit values from a SHAKE256 stream seeded with 32 fresh bytes from the entropy source. The unsplit key is then wiped, and the shares stay in the 64 KB CCM, where the fault handlers know to erase them.

### 6.2 Refresh in two passes, in random order

Before every signature the shares are refreshed with a new uniform $r$ per coefficient, and the order of that work turned out to matter more than its arithmetic. A first loop computes $\hat{s}^{(0)} \leftarrow \hat{s}^{(0)}-r$ for every coefficient, and a separate second loop computes $\hat{s}^{(1)} \leftarrow \hat{s}^{(1)}+r$. Updating both shares of one coefficient back to back would move them through the same registers and bus in consecutive instructions, and on many cores the leakage then depends on the Hamming distance between the two shares, which depends on the secret, the effect Balasch and colleagues measured [18] and Section 10.6 shows in emulation on this code.

Each pass also visits the 256 coefficients of a polynomial in a fresh, uniformly random order, a Fisher-Yates permutation drawn from the masking stream, and the share-wise products during signing are shuffled in the same way, with a random choice of which share goes first. The order changes nothing in the result, because every coefficient is computed from its own inputs only, and the tests check that no shuffle seed changes a signature. Shuffling makes it harder to combine the leakage of the two shares of one coefficient or to average one coefficient over many signatures, without preventing either [19], and the NTTs themselves still run in their fixed order.

## 6.3 Products one share at a time

The two products that touch the secret are computed one share at a time:

$$\boldsymbol{z} = \boldsymbol{y} + \mathrm{NTT}^{-1}\left(\hat{c} \circ \hat{\boldsymbol{s}}_1^{(0)}\right) + \mathrm{NTT}^{-1}\left(\hat{c} \circ \hat{\boldsymbol{s}}_1^{(1)}\right),$$

$$\boldsymbol{r}_0 = \boldsymbol{w}_0 - \mathrm{NTT}^{-1}\left(\hat{c} \circ \hat{\boldsymbol{s}}_2^{(0)}\right) - \mathrm{NTT}^{-1}\left(\hat{c} \circ \hat{\boldsymbol{s}}_2^{(1)}\right).$$

Each term $\mathrm{NTT}^{-1}\left(\hat{c} \circ \hat{\boldsymbol{s}}^{(i)}\right)$ is, by itself, independent of the key. The value becomes secret-dependent only when the second share has been added, and at that point it is $\boldsymbol{z}$ or $\boldsymbol{r}_0$, exactly what the unmasked reference code computes, which is why the masked signer can stay byte-identical to the reference.

## 6.4 What is and is not masked

Table 3 lists the secret-dependent values and their state, and the most important entry in it is an unmasked one. Because $\boldsymbol{z} = \boldsymbol{y} + c\,\boldsymbol{s}_1$ and $\boldsymbol{z}$ is public for an accepted signature, an attacker who learns $\boldsymbol{y}$ through a side channel learns $c\,\boldsymbol{s}_1$ and, with the public $c$, the secret $\boldsymbol{s}_1$ itself. Masking the two products therefore does not make Zirah a first-order secure ML-DSA implementation. It protects the operation that the published attack on this family of chips targets [14], and it is a base on which the rest could be built.

**Table 3. Secret-dependent values in Zirah's signer and whether they are masked.** Source: firmware/src and docs/DESIGN.md in the Zirah repository [23].

| Value | Masked | Comment |
|---|---|---|
| $\boldsymbol{s}_1$, $\boldsymbol{s}_2$ in RAM | Yes | Two shares in CCM, refreshed before every signature in random order |
| $c\,\boldsymbol{s}_1$, $c\,\boldsymbol{s}_2$ products | Yes | Computed share by share in random order |
| $\boldsymbol{z}$ after recombination | No | Public when the signature is accepted; secret-dependent for rejected attempts |
| $\boldsymbol{r}_0$ after recombination | No | Its norm check runs unmasked |
| $\boldsymbol{y}$ from ExpandMask | No | The main gap, see text |
| $K$ and $\rho'$ derivation | No | The hedged `rnd` changes $\rho'$ for every signature |
| $\boldsymbol{t}_0$ and the hint computation | No | Not masked |
| Secret key at rest | No | Packed bytes in flash sector 7, split into shares at every boot |

## 6.5 Memory and parameter sets

Zirah ships at ML-DSA-65, NIST security category 3, and the memory map explains why it stops there. In the DevEBox build examined for this paper, initialised and zeroed data take 75,608 bytes of the 128 KB main SRAM, the masked key and the large work buffers occupy a fixed layout in the 64 KB CCM, and the largest stack frame on the signing path, that of the masked signing function, is 30,456 bytes. ML-DSA-44 also builds and needs less, 52,232 bytes of data and a 22,264-byte frame. ML-DSA-87 does not fit, and its build stops at a compile-time assertion because the CCM layout would exceed 64 KB, so fitting it would mean generating the public matrix $\boldsymbol{A}$ row by row instead of storing it whole and reworking the masking around that, which is real work rather than a configuration change.

## 7. Where the Post-Quantum Key Lives

### 7.1 What the secure element protects

The ATECC608 holds a P-256 private key that it generated itself in slot 0, and for that key it does everything a secure element should. Provisioning writes the slot and key configuration, locks the configuration zone, generates the key, locks the data zone and finally locks slot 0 itself, after which the key can sign external digests but no command can read it. Every signature's classical half comes from it, and its random generator contributes 32 bytes to every key generation, signature and identity proof, mixed into the conditioner and credited with zero entropy (Section 5.4).

It cannot do the same for the post-quantum key, for three separate reasons that do not depend on each other. The chip has no lattice arithmetic, so it cannot compute an ML-DSA signature, and its data zone of 1,208 bytes is split into 16 slots, the largest of which holds 416 bytes [5], about a tenth of the 4,032-byte ML-DSA-65 secret key. Keeping the key there in encrypted form would not help either, because it would have to be decrypted into STM32 RAM before use, and the STM32 would again be the place to attack it. A label such as *secure element inside* is therefore true of the classical key only.

### 7.2 Readout protection on the DevEBox

The ML-DSA-65 key lives, unmasked, in the last 128 KB sector of the STM32's flash, sector 7 on the STM32F407VE, and is split into shares at every boot. At readout protection level 0 a debugger reads that sector like any other memory. On the DevEBox, level 1 was set through the ROM bootloader's Readout Protect command [24], and the firmware, which reads the option bytes at every `INFO` request, reported level 1 in every check afterwards. At level 1 the STM32F407 blocks flash access from a debugger and from the system bootloader, and going back to level 0 erases the flash, key included [21]. No debugger read of the key was attempted on this board, which has no on-board debugger, and published work has defeated readout protection on other STM32 families [22], so level 1 raises the cost of extracting the key without ruling it out. Level 2, which disables the debug interface for good and cannot be undone, was not used.

### 7.3 Key extraction on the earlier board

The weakness that readout protection covers was demonstrated before the DevEBox existed in this project. The firmware first ran on an STM32F4DISCOVERY, whose on-board ST-LINK gives a debugger direct access over SWD. On that board, at readout protection level 0, I copied the ML-DSA-65 key out of the flash sector that holds it, and the copied key produced a valid ML-DSA-65 signature without the board, verifying under the device's public key. The ATECC608 on the same build kept its ECDSA key the whole time, and a copied key cannot produce the ECDSA half, so the forgery covered the post-quantum half only and could not on its own pass as a complete Zirah signature, which needs both halves. Against an attacker who can also break ECDSA, however, the post-quantum half is the whole of the protection, and on that board it rested on nothing at all.

I report this experiment as an account rather than as reproducible evidence. The Discovery was later damaged by an accidental short circuit, and no logs or captures of the extraction are part of this release, so the measurements in Section 10 and the dataset all come from the DevEBox. What the earlier board shows, and what the DevEBox's readout protection only postpones, is the central point of this paper: a device can contain a secure element and still keep its post-quantum key no better than the microcontroller around it. A secure element with lattice arithmetic and enough storage would close that gap, and until such parts are as cheap and easy to buy as the ATECC608, a design like this one has to protect the post-quantum key with the microcontroller's own, weaker means.

# 8. The Host Software

The host software is an extension of the device rather than its centre, and its jobs are to give the board something to sign, to check what comes back and to store the result without touching the original file. One Python library implements those rules and is shared by a command-line tool and a PySide6 desktop application, so both follow exactly the same policy. Because the board signs a digest and never sees a file, it is the host that decides which files can be signed and where each signature travels, and those decisions are what this section describes.

## 8.1 What Zirah signs

Zirah signs any regular file, whatever its format or size, because the signature never depends on understanding the file. The PC reads the file through SHA-256 in 1 MiB pieces, so its size is limited only by the disk, and the board signs a 32-byte digest of a manifest built around that hash (Section 8.2). A firmware image, a compiled program, a database, a video and a contract therefore take exactly the same path, and the format decides only where the signature travels, as Table 4 sets out. For the five formats that have a place readers skip, Zirah writes a signed copy next to the original, for example `CV.signed.pdf`, and removing the signature block from that copy gives back the original bytes exactly, which is what the signature covers. Every other file receives a detached record, `name.ext.zirah.json`, saved beside it.

**Table 4. File formats and where the signature travels.** Source: host/zirah_embed.py and host/zirah_core.py in the Zirah repository [23].

| File type | Recognised by | Where the signature goes |
|---|---|---|
| PDF | A `%PDF-` header at the start | Signed copy: a comment line after the original `%%EOF`, then the original `startxref` offset and `%%EOF` again, so readers find the same cross-reference data |
| Word, Excel and PowerPoint (Office Open XML) | A ZIP header and one of the endings .docx, .docm, .dotx, .dotm, .xlsx, .xlsm, .xltx, .xltm, .pptx, .pptm, .potx, .potm, .ppsx or .ppsm | Signed copy: a stored part `zirah/signature.xml` that no relationship references, so Office applications never load it |
| ZIP archives | A ZIP header and the .zip ending | Signed copy: the archive comment, after any comment already present |
| PNG images | The PNG signature at the start, whatever the ending | Signed copy: an ancillary private chunk `zrAH` before `IEND` |
| JPEG photographs | A JPEG start-of-image marker, whatever the ending | Signed copy: after the end-of-image marker |
| Every other file, including programs and installers, firmware images, source code, text, CSV and JSON, audio and video, disk images, databases, OpenDocument, EPUB, JAR and APK files, older binary Office files and files with no extension | Anything else | Detached record `name.ext.zirah.json` beside the file |

A few rules hold for every format, and they follow from the requirement that a signature must never be ambiguous about which bytes it covers. A file whose structure leaves no safe place for the block falls back to a detached record, which happens with a PDF that lacks a final `startxref`, a ZIP archive with ZIP64 records, a JPEG with data after its end marker or a PNG that does not end in `IEND`. OpenDocument, EPUB, JAR and APK archives and the Open Packaging formats that are not Office files are left detached on purpose, because their readers check the archive more strictly or sign its parts themselves. The host refuses anything that is not a regular file, refuses a file that already carries a Zirah signature, since signing a signed copy again would nest one signature inside another, and refuses a file whose size or modification time changes while it is hashed or between the hash taken before signing and the one taken after. Any file can also receive a detached record on request, and renaming a signed file does not invalidate its signature, because the verifier compares the content hash, the size and the container and reports the signed name.

What a valid Zirah signature asserts is narrow by design. It says that these exact bytes, under the signed name and size and with the times the signing PC reported, were approved by a button press on the device that holds the trusted keys. It does not make a file safe to open or run, and operating systems, PDF readers, Office applications and bootloaders know nothing of Zirah blocks, so a signed program still appears unsigned to Windows code signing and a signed firmware image is not checked by the STM32's bootloader. A Zirah signature is checked with Zirah's own verifier, which needs no board.

The signed copies were checked during development with qpdf, pypdf, Python's zipfile, python-docx, python-pptx, openpyxl, LibreOffice and Pillow, and the host test suite round-trips every format. They were not opened in Microsoft Office or Adobe Acrobat during this work. A copy that is later saved again by an editor no longer carries a valid signature, which is the right result, because it is no longer the file that was signed.

## 8.2 The manifest, the signature record and the verification policy

The manifest is what turns a hash of a file into a statement about that file. It records the file's SHA-256, its name, its size in bytes, the container the signature travels in (`detached`, `pdf`, `ooxml`, `zip`, `png` or `jpeg`), the signing time from the PC's clock, and the file's creation and modification times, all in UTC. It is serialised as JSON with sorted keys, no spaces and ASCII escapes, so that every verifier rebuilds the same bytes, and the board signs the digest of an ASCII tag, a newline byte (0x0A) and those manifest bytes,

$$D = \text{SHA-256}\left(\text{ZIRAH-MANIFEST-v1} \parallel \text{0x0A} \parallel \text{manifest}\right).$$

ML-DSA-65 signs $D$ in the pure mode of FIPS 204 with the context string `zirah-doc-v1` and with hedged randomness, and ECDSA P-256 in the ATECC608 signs the same $D$. The context string is a fixed protocol label that keeps the name it received in the first version of the firmware, and it covers files of every type. Because this is pure ML-DSA applied to a 32-byte message rather than HashML-DSA, a verifier has to use the same mode and context, and the identity proof used during setup signs under a different context, `zirah-attest-v1`, so it can never pass as a file signature. The construction as a whole relies on the collision resistance of SHA-256. The times in the manifest are the signing PC's statements, which the signature fixes but cannot vouch for, since the firmware keeps no clock of its own and an independent time would need a timestamping service.

The signature record, format `zirah-signature-v2`, holds the manifest, the context string, the device's identity and firmware version, both public keys with their fingerprints, both signatures, the device's own timings and a flag that marks simulated results. A signature is accepted only when every one of the following conditions holds.

1. The file's SHA-256 matches the manifest, and so do its size and the container in which the record was found, so a record lifted out of a signed copy into a separate file is refused.
2. Nothing follows a record embedded in a signed copy except the structure Zirah wrote, so data appended later, such as a PDF incremental update, is caught.
3. The context is exactly `zirah-doc-v1`, a value the verifier pins in code and never reads from the file.
4. The ML-DSA-65 signature verifies over $D$, and so does the ECDSA signature.
5. The public keys belong to a trusted device, taken from the PC's trust store or from a key file the signer exported, and once the trusted entry holds an ECDSA key a record without an ECDSA half is rejected, so the classical half cannot be stripped away.
6. The record is not marked as simulated. A missing flag counts as simulated, and the simulator's throwaway keys never enter the trust store.

The result is the AND combination of Section 2.5, so forging a Zirah signature needs both ML-DSA-65 and ECDSA P-256 broken, or both keys stolen. Verification needs no board, and the people who check a Zirah signature install the same program and use its Verify page or the `zirah verify` command.

### 8.3 The desktop application

The desktop application in Figure 8 is a local PySide6 program that opens no network port. It follows the same rules as the command-line tool because both call one host library, and its live trace shows the same block of raw noise samples that the OLED draws, fetched from the device.

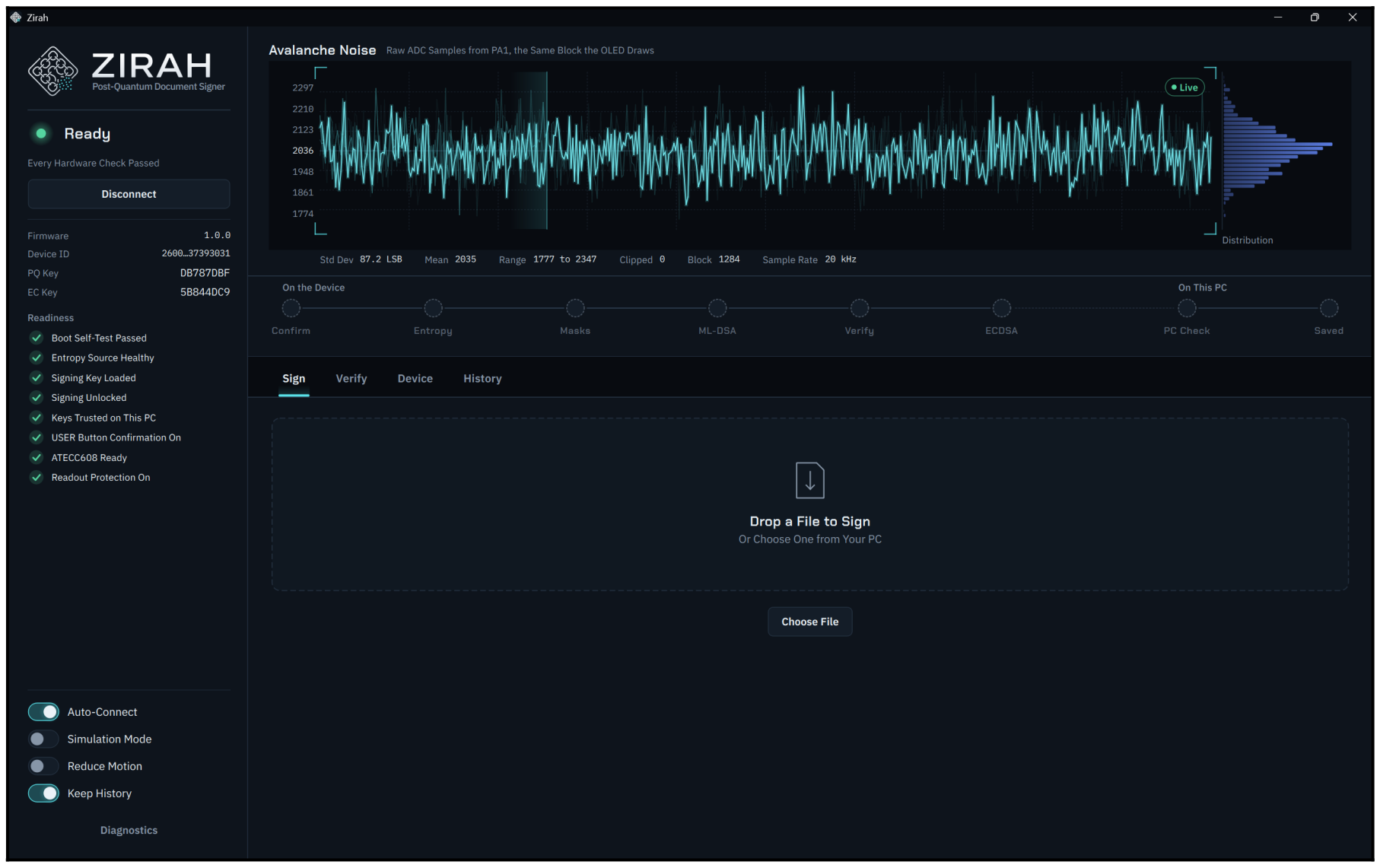

***Fig. 8*** *The desktop app's main screen with the board connected and every hardware check passed. This screenshot was taken with the host build used during the measurement session, whose labels still read Document Signer and PA1, the Discovery board's noise pin; the released host code reads File Signer and takes the pin name, PA4 on the DevEBox, from the board*

## 9. Methodology

### 9.1 Conditions and toolchain

All measurements come from one DevEBox with one Board A and one Board B, taken in Karachi on 5 November 2026 over about two hours, at a room temperature that was not recorded. The board ran firmware 1.0.0, whose self-reported image hash, the first 16 bytes of SHAKE256 over the flash image, matched the `zirah-devebox.bin` on the PC, and the firmware test suite rebuilt that image byte for byte from the source. The PC ran Zirah's host software 1.0.0 on Windows. The STM32's internal temperature sensor read between 35 and 48 °C, and since its single readings are noisy and the die runs warmer than the air around it, those values describe the chip rather than the room. The software test suites ran on Ubuntu 24.04 with Python 3.13.18, GCC 13.3.0, arm-none-eabi-gcc 13.2.1, QEMU 8.2.2 and PySide6 6.8.2.1.

### 9.2 Entropy capture and the NIST SP 800-90B assessment

The raw samples were captured with `zirah capture`, which asks the firmware for unconditioned ADC values, and assessed with NIST's own SP 800-90B tool. A first attempt to transfer 1,000,000 samples in a single request stalled on the PC side after the board had finished sampling, with the host waiting for bytes that never arrived, so the data were captured again in 20 requests of 50,000 samples each, about 11.5 s of transfer per piece, with a CRC-32 over every piece checked on the PC. The firmware records raw data in blocks of 2,048 samples and pauses between blocks to send them, so the samples are contiguous only inside a block, and each piece holds 24 full blocks and one partial block of 848 samples. The 20 pieces were concatenated in order into one file of 1,000,000 samples, from which the low bytes formed the 8-bit symbols and bit 0 formed a separate binary sequence.

Both files went through the tool's `ea_iid` and `ea_non_iid` programs [8] at commit 87c104d0 (May 2026), built on Linux, with the commands recorded in the dataset [6]. The tool reads a file as one sequence, so the block joins are part of what it assessed. Autocorrelation and spectra were computed separately inside the contiguous blocks, where the joins cannot distort them.

## 9.3 Correctness checks

Masking must not change a single output bit, so correctness was checked against the reference rather than against itself. The `mldsa` suite compiles the firmware's own signer for a PC, under AddressSanitizer and UBSan, and checks that key generation matches the reference `crypto_sign_keypair` byte for byte, that masked signatures match the reference signer for the same key, message, context and `rnd` over several keys, message lengths, contexts and refreshes, that the shuffle is a uniform permutation that never changes a signature, and that the on-device verifier agrees with the reference verifier on valid and altered inputs. The `qemu` suite builds the same program for the Cortex-M4 with the assembly NTT and runs it on QEMU's `mps2-an386` machine, comparing a hash of every key and signature it produces with the hash from the PC build, and a further set of checks compares the assembly NTT with the C NTT modulo $q$ for inputs from every range the signer uses. The assembly folds its final scaling into the last layer, so its outputs equal the C outputs modulo $q$ rather than as integers, which is harmless only because every later step reduces first, and those checks cover that point. The firmware also repeats a short known-answer test at every power-up (SHAKE256, key generation from a fixed seed, a masked signature and its verification) against fingerprints computed with dilithium-py rather than with the firmware, and the `kat` suite recomputes those four vectors.

## 9.4 Leakage and timing checks

During development the refresh-and-multiply kernel was compiled for Thumb-1 and run through ELMO, which emulates the machine code and outputs a modelled power value per instruction [27], under two sets of model coefficients, one for the Cortex-M0 and one for the Cortex-M3. A fixed-versus-random Welch $t$-test [28] with 5,000 traces per group was computed for four variants of the kernel: unmasked, naively masked with both shares of a coefficient handled back to back, masked in two passes, and masked in two passes with shuffling. Values of $|t|$ above 4.5 count as detected leakage. No physical trace was recorded on the STM32F407, which is a Cortex-M4 with a different pipeline from either model.

Running time is a separate channel, and the `timing` suite follows dudect [29]. For six kernels that handle key material, a constant-time compare, the share-by-challenge product, both NTTs, the reduction step and the whole mask refresh, it times each call with the x86 time-stamp counter while the secret input is, at random, either a fixed value or a fresh one, and compares the two groups with Welch's $t$-test, also after cropping the slowest calls. A deliberately leaky compare that stops at the first differing byte serves as the positive control. For the Cortex-M4, whose divider takes a time that depends on its operands, the firmware suite disassembles the 12 objects that handle the key and fails if any contains a `udiv` or `sdiv` instruction.

## 9.5 Software test suites

Zirah's test runner has ten suites, listed in Table 5, and each one reports SKIP with a reason rather than passing when its tools are missing. All ten ran on 5 November 2026 after the host fixes made during the hardware session, and Table 5 gives the final full run on the released code, in which all ten passed. The host suite has 64 tests, and the app suite drives the real desktop window offscreen against the simulated device, which can never produce a signature that verifies as a hardware signature.

**Table 5. Test suites run for this paper, with the final run's results. All passed.** Source: tests/run_tests.py in the Zirah repository [23].

| Suite | What it checks | Result |
|---|---|---|
| `oled` | The firmware's OLED code built for the PC; every screen rendered, any clipped text fails | 43 screens, none clipped |
| `mldsa` | The firmware's signer against the reference, under ASan and UBSan | 1,976 checks |
| `entropy` | Health tests, statistics, the two-source correlation and the jitter checks on synthetic sources | 2,234 checks |
| `timing` | dudect on six key-handling kernels, with a leaky control | Largest $\lvert t \rvert$ 2.11; control 1,829 |
| `kat` | Boot self-test vectors recomputed with dilithium-py | 4 vectors |
| `style` | Title Case for every OLED and desktop string | 679 strings |
| `host` | Protocol, trust, verification, signed copies in every format, the serial flasher | 64 tests |
| `app` | The desktop app through sign and verify, start and quit, drawing time | 0 failures |
| `qemu` | The assembly NTT and the `mldsa` suite on an emulated Cortex-M4 | 3,001 and 1,976 checks |
| `firmware` | ARM builds for both boards, zero warnings, no divide instruction in key code, images equal to the shipped ones | Both identical |

# 10. Results

## 10.1 Readiness and health checks

`zirah validate` passed all 12 readiness checks and reported the state Ready. The checks covered firmware and protocol compatibility, the boot self-test, the health of the entropy source, the loaded and masked key, unlocked signing, trusted keys, the OLED, button confirmation, the ATECC608, the MPU and stack canary, readout protection, and the identity of the firmware with the PC's copy. Its start-up test passed with the bias supply at 13.96 V, and in 1,024 symbols the longest run of one symbol was 2 (cutoff 21) and the largest adaptive-proportion count was 2 (cutoff 311). A 4,096-sample block had a mean of 2,024, a standard deviation of 94.8 ADC steps, a range of 1,698 to 2,367 with nothing clipped, all 256 low-byte values present and a lag-1 autocorrelation of 0.091. The identity proof verified on the PC, and an `INFO` request later in the session showed 103,936 symbols through the continuous tests since power-up and no failure.

## 10.2 Raw noise statistics, correlation and spectrum

Every one of the 20 capture pieces arrived with a matching CRC-32, no health-test failure occurred during any piece, and the bias supply read between 13.57 and 14.25 V at the start of each. Over all 1,000,000 samples the mean was 2,052.0 and the standard deviation 95.5 ADC steps, with a minimum of 1,659, a maximum of 2,705 and no clipping (Figure 9). The per-piece means stayed between 2,050.2 and 2,052.7 and the per-piece standard deviations between 94.9 and 96.2, so the source was stable over the four minutes of capture.

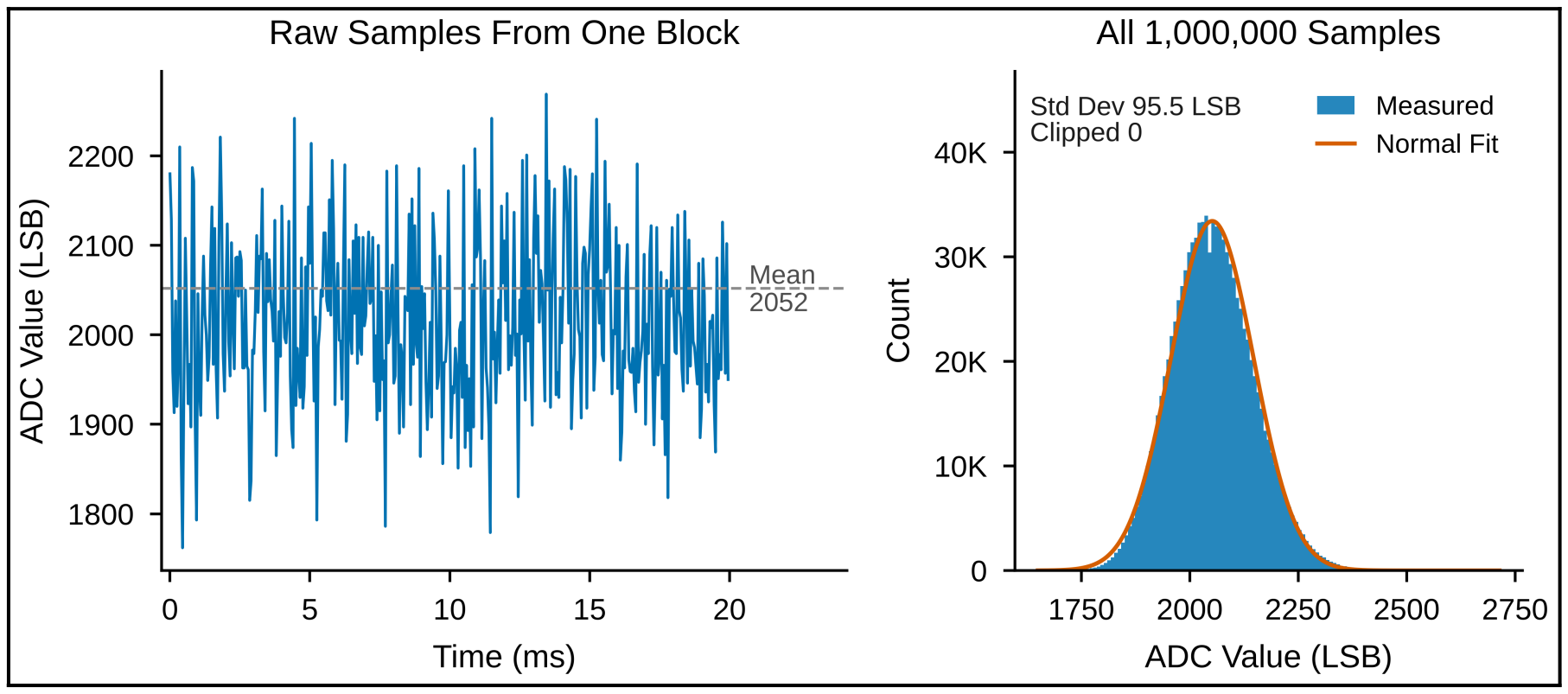

***Fig. 9*** *Left, the first 400 samples of the capture, 20 ms at 20 kHz. Right, the histogram of all 1,000,000 samples with a normal curve of the same mean and standard deviation*

Computed inside the contiguous blocks, the autocorrelation is 0.096 at lag 1, 0.056 at lag 10 and 0.028 at lag 40 (Figure 10), far outside the band of ±0.002 expected for independent samples, so neighbouring samples are correlated and the decay is slow. The averaged spectrum carries distinct lines at 115.7, 231.3, 346.7, 462.3 and 578.0 Hz, harmonics of about 115.6 Hz, which together hold about 5 % of the variance. Another 4 % lies below 50 Hz, and from 1.2 to 10 kHz the spectrum is flat to within about 1 dB. The lines are not multiples of the 50 Hz mains frequency. The boost converter and digital activity on shared supply or ground paths are candidates, but no oscilloscope was available to separate them, so the source remains unidentified. The firmware has a check for exactly this kind of common signal, a correlation test between two noise sources that faults above 0.30, but this build has only one source, and a signal holding 5 % of the variance would stay below that limit anyway.

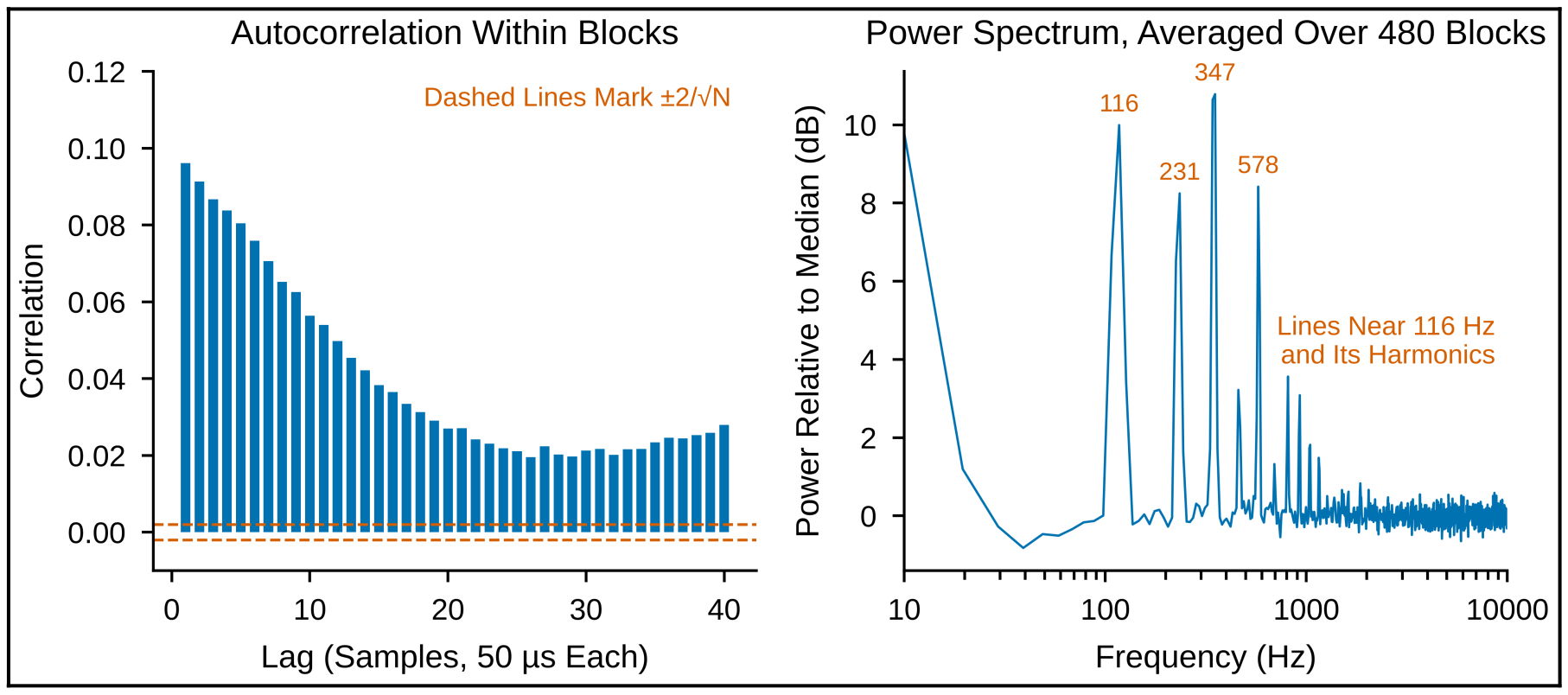

***Fig. 10*** *Left, the autocorrelation computed inside the contiguous 2,048-sample blocks, with dashed lines at ±2/√N. Right, the power spectrum averaged over the 480 full blocks with a Hann window, relative to its median, with the labelled lines at harmonics of about 115.6 Hz*

## 10.3 NIST SP 800-90B assessment

The NIST tool rejected the IID assumption for these samples, and Table 6 shows where. The chi-square tests and the longest-repeated-substring test passed, but 3 of the 19 permutation tests failed: the number of runs about the median and the covariance at lags 1 and 2. In each failed test the original data ranked beyond all but at most one of the 10,000 shuffled copies, in the direction of positive correlation between neighbouring samples, which agrees with the autocorrelation in Section 10.2. The data are therefore not IID, and the non-IID estimates are the ones that apply.

**Table 6. SP 800-90B IID-track tests on the 1,000,000 8-bit symbols.** Each permutation test compares the original statistic with 10,000 shuffled copies. Source: nist/ea_iid_sym8.txt in the Zirah dataset [6].

| **Test** | **Result** | **Outcome** |
|---|---|---|
| Chi-square independence | Score 63,922.5 on 63,886 degrees of freedom, p = 0.459 | Passed |
| Chi-square goodness of fit | Score 2,294.8 on 2,295 degrees of freedom, p = 0.497 | Passed |
| Longest repeated substring | Length 4 | Passed |
| 16 of the 19 permutation tests | Original within the range of the shuffled copies | Passed |
| Number of runs about the median | Original below all 10,000 shuffled copies | Failed |
| Covariance at lag 1 | Original above all 10,000 shuffled copies | Failed |
| Covariance at lag 2 | Original above 9,999 of the 10,000 shuffled copies | Failed |

Table 7 and Figure 11 give the non-IID estimates. For the 8-bit symbols the tool reports $H_{\text{original}} = 7.059$, the $t$-tuple estimate, and $H_{\text{bitstring}} = 0.862$, the compression estimate on the 8,000,000 bits of those symbols, and the assessed min-entropy is

$$H = \min\left(H_{\text{original}}, 8 \times H_{\text{bitstring}}\right) = \min\left(7.059, 6.898\right) = 6.898 \text{ bits per symbol}.$$

Bit 0 on its own assessed at 0.784 bits, also set by the compression estimate, with about 52.0 % of its values equal to one. The firmware’s own quick estimate in the validation run, 6.42 bits per symbol from one 4,096-sample block, is lower mainly because the confidence bound on a small sample is wider, and it is a sanity check rather than an SP 800-90B assessment.

**Table 7. SP 800-90B non-IID min-entropy estimates for 1,000,000 samples.** The middle column applies the estimators to the 8,000,000 bits of the 8-bit symbols; n/a marks estimators that run on bit strings only. Source: nist/ea_non_iid_sym8.txt and nist/ea_non_iid_bit0.txt in the Zirah dataset [6].

| **Estimator** | **8-bit symbols** | **Their bits** | **Bit 0 only** |
|---|---|---|---|
| Most common value | 7.249 | 0.994 | 0.941 |
| Collision | n/a | 1.000 | 0.945 |
| Markov | n/a | 0.993 | 0.947 |
| Compression | n/a | 0.862 | 0.784 |
| $t$-tuple | 7.059 | 0.926 | 0.899 |
| Longest repeated substring | 7.895 | 0.988 | 0.985 |
| Multi most common in window | 7.511 | 0.992 | 0.948 |
| Lag prediction | 7.905 | 0.987 | 0.994 |
| Multi Markov model with counting | 7.597 | 0.972 | 0.941 |
| LZ78Y prediction | 7.597 | 0.994 | 0.941 |
| Minimum | 7.059 | 0.862 | 0.784 |

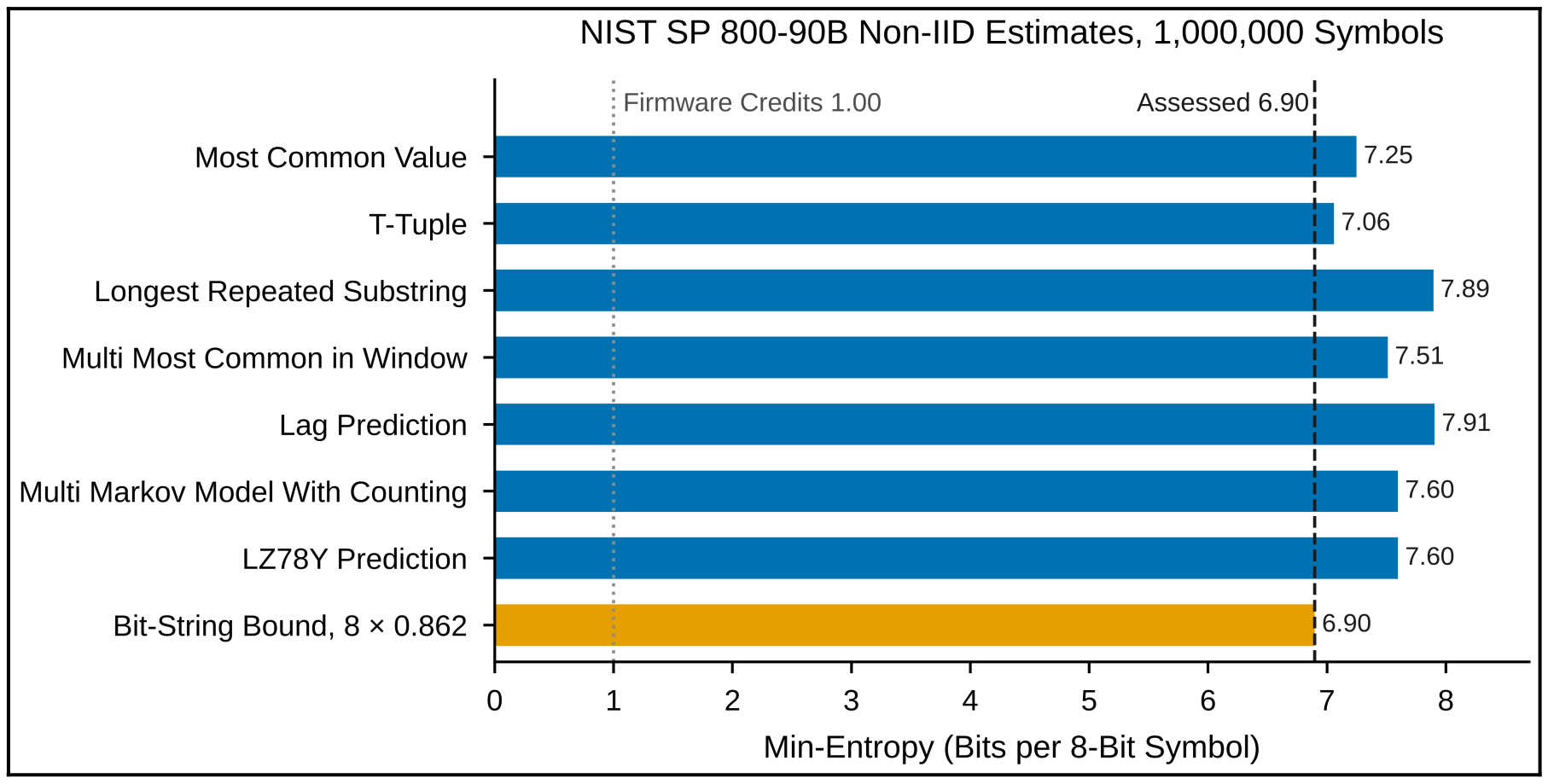

***Fig. 11*** *The non-IID estimates for the 8-bit symbols. The assessed value is set by the bit-string bound, eight times the compression estimate on the bits, and lies far above the 1.00 bit per symbol that the firmware credits*

The firmware credits 1.00 bit per symbol, so each 32-byte output block absorbs 512 symbols credited with 512 bits of entropy, while by this assessment those symbols carry about 3,530 bits. That margin is the reason I left the credited value alone. The assessment covers one board, one transistor, one session and one supply, the restart tests of SP 800-90B Section 3.1.4 were not run, and the spectral lines and slow correlation are unexplained, so a value assessed under those conditions is evidence about this source rather than a number to build into the firmware.

## 10.4 Correctness of the masked signer

The masked signer produced the same bytes as the reference in every case tested. All 1,976 checks of the `mldsa` suite passed natively under AddressSanitizer and UBSan, the same 1,976 passed on the emulated Cortex-M4 with the assembly NTT, and the hash over every key and signature from the emulated run equalled the hash from the PC build. The 3,001 comparisons of the assembly NTT with the C NTT modulo $q$ all agreed, and the four boot-time vectors recomputed with dilithium-py matched the firmware's. In use, every signature is verified once on the device and once more on the PC, so a fault that changed a signature would have to change it in a way that still verifies twice.

## 10.5 Performance on the board

Table 8 gives the timings that `zirah validate` collected on the board for 20 signatures with a throwaway key, timed with the CPU cycle counter while the ATECC608 generator, the STM32 generator and the clock jitter were all mixed in. Rejection sampling dominates the spread, from one attempt to 24 at about 27 ms per attempt on average, which is why the median and the mean differ so much. The same run measured key generation at 79.7 ms, the boot-time load and split of the key at 58.9 ms, and SHAKE256 over 1 KB at 271,922 cycles. The assembly NTT took 8,879 cycles for the forward and 9,455 for the inverse transform, against 31,295 and 31,932 for the reference C code on the same chip, a speed-up of a little over three times.

**Table 8. The masked signing path on the board, 20 signatures with a throwaway key, NTT in assembly, 168 MHz.** Source: validation/validation-20261105-101549.json in the Zirah dataset [6].

| Stage | Min | Median | Max | Mean |
|---|---|---|---|---|
| Entropy for 32 bytes (ms) | 36.5 | 36.5 | 36.5 | 36.5 |
| Mask refresh (ms) | 63.8 | 64.2 | 64.2 | 64.1 |
| Masked signature (ms) | 57.8 | 157.4 | 519.8 | 217.5 |
| Rejection attempts | 1 | 5 | 24 | 8 |
| On-device verification (ms) | 39.9 | 39.9 | 39.9 | 39.9 |

Table 9 gives the device’s own timings for two real signatures with the device key, read from the signature records and from the app. Without the wait for the button, the device spent between 0.67 and 0.76 s on each. The fresh start-up test and the mask refresh with its hedging randomness together take about 0.4 s, more than the signature itself, and both are deliberate costs of failing closed. The ATECC608 took 94.6 ms for one ECDSA signature and 130.4 ms for the other, and I did not investigate the reason for the difference.

**Table 9. Device time for two signatures with the device key.** The first is a 396 KB PDF signed with a detached record, the second a DOCX file signed into a signed copy. Source: the device timings stored in the two signature records, as the desktop app reports them.

| Stage | PDF (ms) | DOCX (ms) |
|---|---|---|
| Start-up health test | 177.9 | 177.4 |
| Mask refresh and hedging randomness | 225.5 | 225.4 |
| Masked ML-DSA-65 signature | 135.5 | 185.8 |
| Rejection attempts | 2 | 4 |
| On-device verification | 39.9 | 39.9 |
| ECDSA in the ATECC608 | 94.6 | 130.4 |
| Total | 673.5 | 758.9 |

## 10.6 Emulated leakage and timing

Table 10 and Figure 12 give the emulated leakage results, and they confirm the reason for the two-pass refresh. The unmasked kernel leaks heavily under both ELMO models. Naive masking, with both shares of a coefficient handled back to back, still leaks, strongly under the Cortex-M3 model and barely under the Cortex-M0 model, while both two-pass variants stay below the threshold of 4.5 under both models. Shuffling does not change these numbers, as expected, because a fixed-versus-random test looks for first-order leakage, which the masking already removes, and the attacks that shuffling slows down, which combine several points or average one coefficient over many traces, are not what this test measures. The very different verdicts the two models give for the naive variant are themselves a reason to repeat the test on the real chip.

**Table 10. Largest |t| from fixed-versus-random TVLA on ELMO traces of the refresh-and-multiply kernel, 5,000 traces per group.** Values above 4.5 count as detected leakage; the results are emulated, not measured. Source: the ELMO evaluation in docs/DESIGN.md in the Zirah repository [23].

| ELMO model | Unmasked | Naive masking | Two-pass | Two-pass and shuffled |
|---|---|---|---|---|
| Cortex-M3 coefficients | 119.37 | 56.41 | 3.65 | 3.65 |
| Cortex-M0 coefficients | 127.43 | 5.12 | 3.44 | 3.34 |

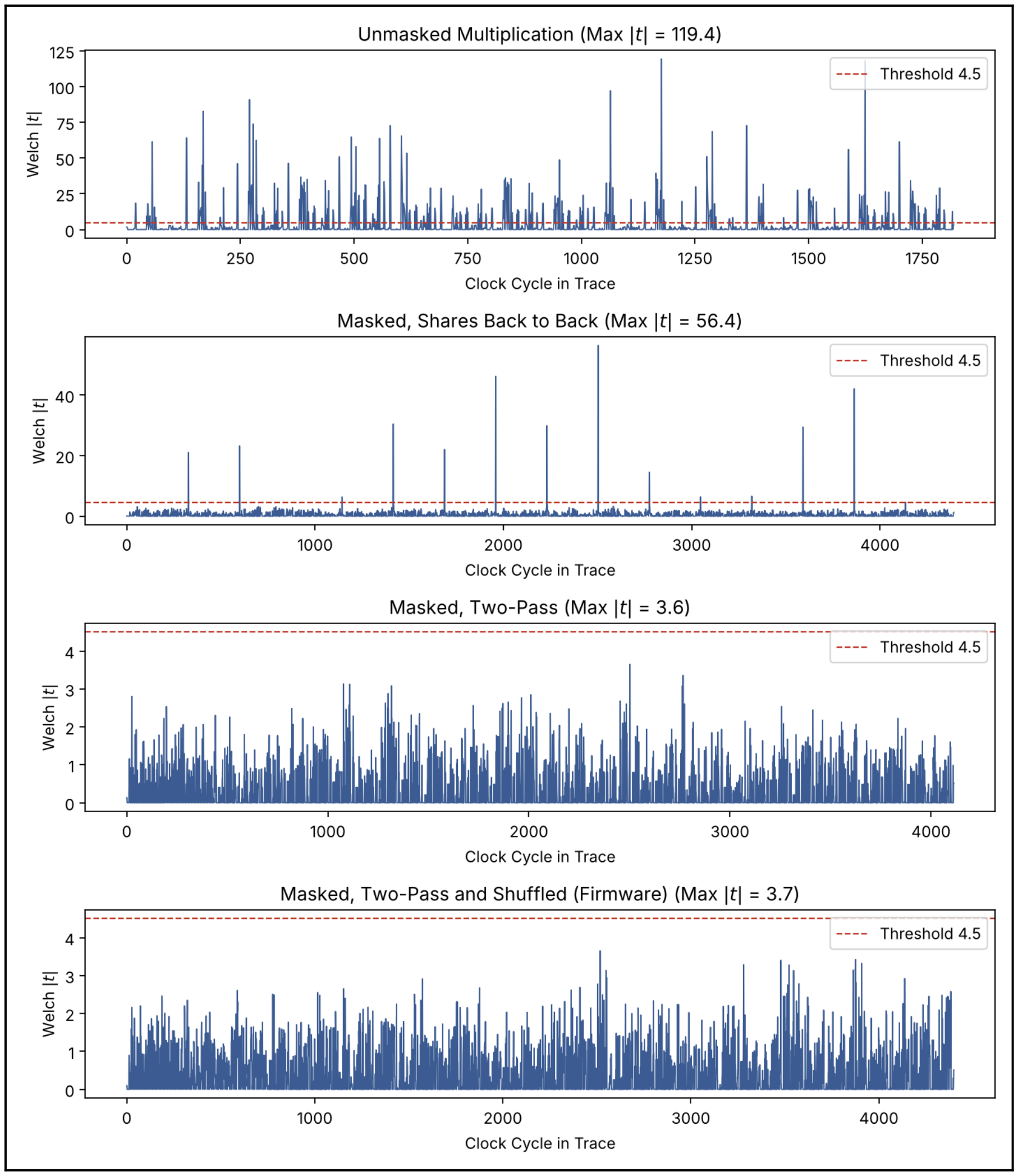

***Fig. 12*** *Welch |t| per emulated clock cycle under ELMO's Cortex-M3 model for the four kernel variants, with the 4.5 threshold dashed. The firmware uses the last variant. This plot is kept at the resolution it was made, because the trace data behind it are not part of this release*

The timing results point the same way on the PC. In six full runs of the `timing` suite on 5 November 2026, the largest $|t|$ over the six key-handling kernels stayed between 1.52 and 3.40, against dudect's leak threshold of 10, while the deliberately leaky compare reached $|t|$ between 615 and 1,829. This is a result about the C code compiled for a PC, where a secret-dependent branch or table lookup would show, and none of the 12 key-handling objects in the ARM build contained a divide instruction. No timing was measured on the board.

## 10.7 Functional tests

The device generated its ML-DSA-65 key with a pairwise consistency test, giving the fingerprint DB787DBF, and the PC pinned it, together with the ATECC608's key 5B844DC9, after the fingerprint was typed from the OLED. A DOCX file was signed into a signed copy and a PDF into a detached record, and Zirah verified both as Valid Hybrid Signature (Figure 13). A copy of the PDF with three bytes appended was rejected: Zirah's verifier, given the signer's exported key file, reported Invalid Signature, because the content hash no longer matched and neither signature verified over the rebuilt manifest.

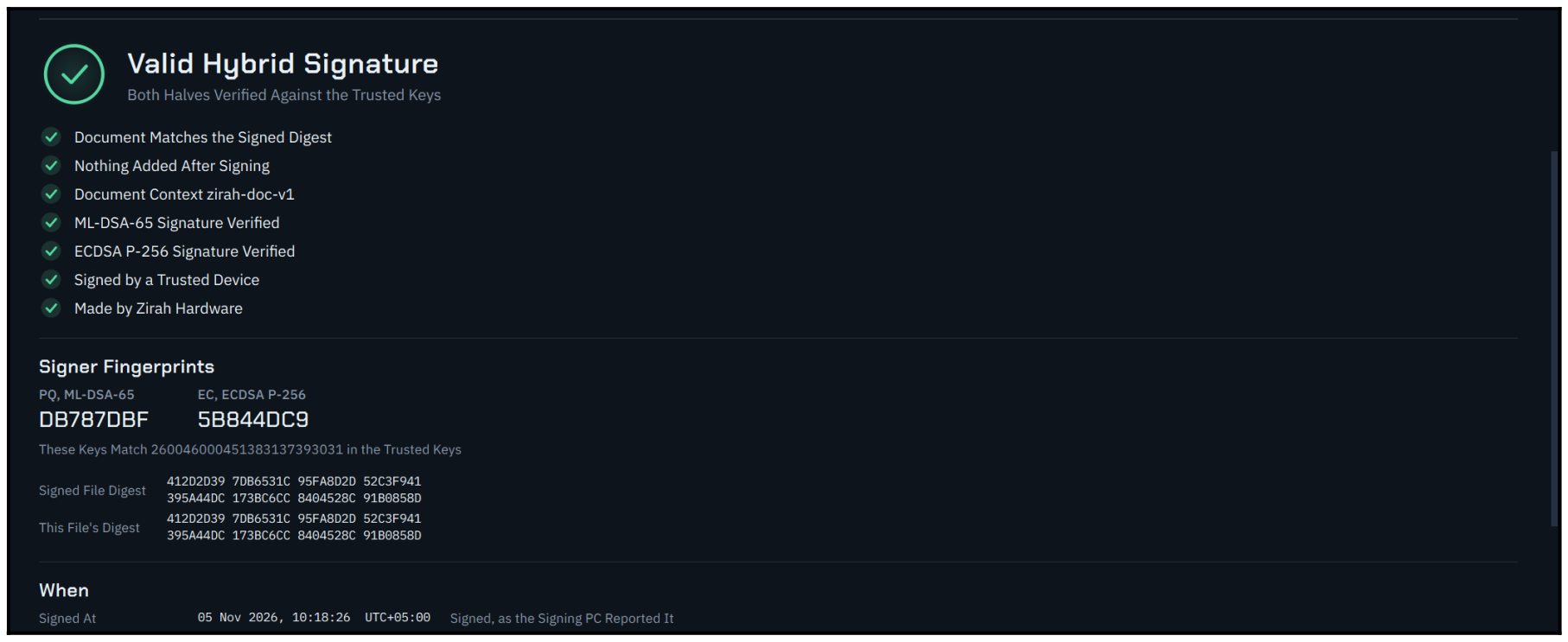

***Fig. 13*** *The Verify page for the signed DOCX copy, cropped to the report: every check passed, the signed digest and the file's digest agree, and the signer's fingerprints are shown in large type for comparison with published ones. In this host build the first and third check names read Document, which the released code names File Matches the Signed Digest and File-Signing Context*

As a check outside Zirah's own code, a short script rebuilt the manifest digest from the PDF's signature record and verified both halves with dilithium-py 1.5.1 and pyca/cryptography 50.0.1. Both halves verified under the context `zirah-doc-v1`, the ML-DSA-65 signature failed under `zirah-attest-v1`, as it should, and the fingerprints recomputed from the public keys matched the ones the OLED had shown. Because the host uses the same two libraries, this check is independent of the firmware and of Zirah's host code but not of those libraries.

## 11. Discussion

The measurements support a plain reading of what a secure element buys in a hybrid design, and I think the reading is less comfortable than the marketing of such parts suggests. The ATECC608 did its job for the classical key, which moved between boards inside its chip and was never exposed, while the post-quantum key had to be generated again on the new board and lived in ordinary flash on both. On the Discovery, a debugger read that key in the most ordinary way, and on the DevEBox only readout protection level 1 stands in front of it. A hybrid signature is meant to stay secure if either algorithm falls, but in a device of this kind its security against a quantum attacker reduces to the microcontroller's readout protection, which published work has already bypassed on related parts. The secure element is not useless here, since it still forces an attacker without a quantum computer to break into the chip for the classical half, yet a product that advertises a secure element and a post-quantum signature in the same sentence invites a conclusion that the hardware does not support.

The masking result is narrower than it may first appear. The two-pass refresh removed the first-order leakage that naive masking left in emulation, and the signer stayed byte-identical to the reference while doing so, which shows that the engineering cost of masking the products is modest. The unmasked $\boldsymbol{y}$, however, leaves a direct path to $\boldsymbol{s}_1$, and an attacker who can recover $\boldsymbol{y}$ has no need to attack the masked products at all. I read the result as evidence that partial masking is cheap and worth doing as a first step, not as evidence that the device resists power analysis.

The entropy results invite the opposite temptation, which is to trust the assessed figure too much. An assessed 6.898 bits per symbol against 1 bit credited is a wide margin, and it would be easy to raise the credit and sign faster. The failed IID tests, the slow correlation and the unexplained lines near 116 Hz all argue against that, because each of them is a sign of structure that a different board, a different supply or a hotter room could make worse. Keeping the credit low costs about 26 ms per output block and nothing else, and on this evidence it is the right trade.

The human confirmation step deserves the least confidence of all. It stops silent signing, and during the session it did exactly that, but a 32-bit prefix shown on a small screen is a check against mistakes rather than a check against a determined attacker who controls the PC. A device that wanted to resist such an attacker would have to show more of the digest, or show a description of the file computed on the device, which in turn would need the device to see the file. Zirah deliberately keeps files on the PC, and the price of that choice is a confirmation step of limited strength.

## 12. Limitations

The masking is partial. Only the products with $\boldsymbol{s}_1$ and $\boldsymbol{s}_2$ are masked, $\boldsymbol{y}$ and its generation are not, and an attacker who recovers $\boldsymbol{y}$ recovers $\boldsymbol{s}_1$. The leakage evidence comes from an emulator of Cortex-M0 and Cortex-M3 cores rather than from the Cortex-M4 in the device, and the timing evidence comes from an x86 build, so no power, electromagnetic or fault measurement was made on the board and nothing here supports a claim that Zirah resists side-channel or fault attacks.

The post-quantum key is protected at rest only by readout protection level 1, which has been defeated on other STM32 families and which was not tested against an attacker in this work. The key extraction reported in Section 7.3 was made on an earlier board that no longer works, and no logs of it are part of this release, so it stands as an account of what happened rather than as reproducible evidence.

The entropy results describe one board, one transistor, one session and one supply. The restart tests were not run, the data were captured in blocks with gaps between them, the samples are correlated, and about 5 % of the variance sits in spectral lines of unknown origin. The boost converter behaved outside its expected range for reasons I did not find, the extra inputs were not assessed, and there is no DRBG after the conditioner. The signing and file times are statements by the signing PC, so a valid signature shows that the signed content, name, size and times have not changed since signing without showing that the times were true.

The signed copies were checked with open-source readers and LibreOffice but not with Microsoft Office or Adobe Acrobat, and the screenshots in Figures 5, 8 and 13 come from the host build used during the session, whose labels predate the released wording. Zirah is a single prototype with no formal certification, no independent audit and no FIPS 140-3 or SP 800-90B validation.

## 13. Related Work

The closest prior work is the line of masked Dilithium implementations, and Zirah sits well below it in protection while differing from it in setting. Migliore and colleagues [15], Azouaoui and colleagues [16] and Coron and colleagues [17] mask the whole signing procedure, including the masking vector and the rejection checks, at first or higher order, and evaluate the result on real or modelled hardware. Zirah masks only the two products that the published attack of Chen and colleagues [14] targets, and it adds the two-pass refresh that the work of Balasch and colleagues [18] motivates. What it contributes in exchange is a complete device around the signer, with key storage, confirmation, verification and an entropy source, so that the remaining weaknesses can be seen in place rather than in isolation.

On performance, the reference points are the Cortex-M4 implementations of Greconici and colleagues [11] and the pqm4 benchmarks [12]. Zirah uses their assembly NTT and reports cycle counts for it on the real chip, but its signing times are dominated by deliberate costs that a benchmark does not carry, a fresh health test, a mask refresh and an on-device verification for every signature, so its numbers are not comparable with benchmark figures and are not meant to be.

The secure-element argument connects to work on readout protection, where Obermaier and Tatschner showed that the protection on STM32F0 devices can be defeated [22], and to the transition guidance that makes hybrid signatures attractive in the first place [1, 3]. In the literature I reviewed, I did not find the specific point that a classical secure element leaves the post-quantum half of a hybrid signature to the microcontroller demonstrated on a complete, low-cost signing device with its key extraction reported, and that demonstration, more than any single number, is what this paper adds.

## 14. Ethical and Dual-Use Considerations

This paper describes a defensive prototype, and the one attack it reports, reading a key out of an unprotected microcontroller with a debugger, uses the same debug access that is a routine part of STM32 development. The extraction was carried out on my own device, with a key I generated for it, and no third-party device or key was involved. The firmware, the host software, the circuit and the measurements are released so that others can check every claim and reuse the parts that work, and the released code contains no tool for extracting keys from other devices. Publishing the gap between a secure element and a post-quantum key seems worthwhile because the gap exists whether or not anyone writes it down, and a designer who knows about it can choose a part that closes it or at least avoid overstating what a product protects. The noise captures contain no personal data.

## 15. Conclusion and Future Work

Zirah shows that an STM32F407 board, a discrete avalanche noise source and an ATECC608 are enough to make hybrid ML-DSA-65 and ECDSA P-256 signatures for files of any type, with a human confirmation step, in well under a second of device time per signature, and that the masked signer can stay byte-identical to the reference implementation while it does so. The measurements also show where such a design stops. The secure element protects the classical key and cannot protect the post-quantum one, which ends up behind the microcontroller's readout protection and which a debugger read out of the earlier board, the masking covers the secret-key products but not the masking vector, and the noise source, while assessed at almost seven bits per byte, carries correlations and spectral lines that a one-board, one-session study cannot explain.

The next steps follow from those limits, and I list them roughly in the order I expect to reach them. The first is physical power and electromagnetic measurements on the STM32F407 itself, which would test the emulated leakage results where they matter. The second is masking $\boldsymbol{y}$ and the rejection checks, so that the signer approaches a first-order secure implementation. The third is a broader entropy study, with restart tests, captures at several temperatures and an oscilloscope to find the source of the lines near 116 Hz, together with the second noise source that the firmware already supports. Further out, generating $\boldsymbol{A}$ on the fly would let ML-DSA-87 fit, and a secure element with lattice arithmetic, once one is as easy to buy as the ATECC608, would let the post-quantum key move off the microcontroller altogether.

## Statements and Declarations

### CRediT Authorship Contribution Statement

Muhammad Shaheer Bin Junaid: Conceptualisation, Methodology, Software, Validation, Formal analysis, Investigation, Resources, Data curation, Writing (original draft), Writing (review and editing), Visualisation, Project administration.

### Declaration of Competing Interest

The author declares that he has no known competing financial interests or personal relationships that could have appeared to influence the work reported in this paper. This research received no specific grant from any funding agency in the public, commercial, or not-for-profit sectors.

### Data and Code Availability

The firmware and host source code, the prebuilt firmware images and their hashes, the circuit diagrams, schematics and board layouts, the interactive build guide, the 20 raw capture files with their metadata, the outputs of NIST's assessment tool, the validation report and the device's public keys are archived on Zenodo at https://doi.org/10.5281/zenodo.23982873 [6], with the development repository at https://github.com/mshaheerjunaid/Zirah [23]. Every table in this paper names its source, and all of those files except the two signature records behind Table 9 are part of the archive, so the numbers can be checked against it. Zirah's own code is released under the MIT licence and the dataset under CC BY 4.0, while the vendored ML-DSA reference code, pqm4's assembly NTT and the other third-party components keep their own licences, which the archive lists.